\documentclass[10pt]{iopart}
\usepackage{graphicx}
\usepackage{color}
\usepackage[english]{babel}
\usepackage{cite}
\usepackage{upgreek}
\usepackage{ulem}

\usepackage{siunitx}

\usepackage[colorlinks=true,linkcolor=black,anchorcolor=blue,citecolor=red,filecolor=green,menucolor=blue,runcolor=blue,urlcolor=blue]{hyperref}

\expandafter\let\csname equation*\endcsname\relax
\expandafter\let\csname endequation*\endcsname\relax
\usepackage{amsmath}
\usepackage{relsize}
\usepackage[colorinlistoftodos]{todonotes}

\newcommand{\ds}{\displaystyle}

\newcommand{\cc}{\mathrm{c}}

\begin{document}

\newcommand{\ud}{\ensuremath{\mathrm{d}}}

\title[Calculate forces and torques on Kibble coil] {Calculation of magnetic forces and torques on the Kibble coil}

\author{Stephan Schlamminger$^1$, Lorenz Keck$^1$, Frank Seifert$^1$, Leon S. Chao$^1$, Darine Haddad$^1$, Shisong Li$^2$}

\date{September 2022}
\address{$^1$ National Institute of Standards and Technology,
100 Bureau Drive, Gaithersburg, MD, USA}
\address{$^2$ Department of Electrical Engineering, Tsinghua University, Beijing 100084, China\\}

\begin{abstract}
Analytically the force acting on a current-carrying coil in a magnetic field can be calculated in two ways. First, a line integral can be conducted along the coil's wire, summing up the differential force contributions. Each contribution results from a cross-product of the corresponding differential line segment with the magnetic flux density at that location. Alternatively, the coil's energy in the field is given as a product of three factors, the number of turns, the current, and the flux threading the coil. The energy can then be obtained by executing a surface integral over the coil's open surface using the scalar product of the differential surface element with the magnetic flux density as its integrand. The force on the coil is the negative derivative of the energy with respect to the appropriate coordinate.
For yoke-based Kibble balances, the latter method is much simpler since most of the flux is contained in the inner yoke of the magnet and can be written as a simple equation. Here, we use this method to provide simple equations and their results for finding the torques and forces that act on a coil in a yoke-based magnet system. We further introduce a straightforward method that allows the calculation of the position and orientation difference between the coil and the magnet from three measurements.

\end{abstract}
\maketitle
\ioptwocol
\section{Introduction}
\label{sec1}

At the time of this writing (May 2022), the  revision of the international system of units (SI) occurred three years ago. Since then, Kibble balances~\cite{Robinson16}, together with the X-ray crystal density (XRCD) method~\cite{Bartl_2017}, have successfully contributed to the mass scale for the world~\cite{Stock2020}. 

In the Kibble balance, a coil with $N$  turns of wire  is immersed in a magnetic field. A vertical force $F_z$ is generated by energizing the coil with current $I$.  It is
\begin{equation}
F_z = -NI \frac{\partial\Phi}{\partial z},
\label{eq:sec1:Fz}
\end{equation}
where $\Phi$ is the magnetic flux through the coil and $z$ the coil position along vertical. The force is compared to the weight $mg$ of mass $m$, and $g$ is the local gravitational acceleration.  As is indicated by the sign in equation~(\ref{eq:sec1:Fz}) and for $I>0$, the force is in the negative direction of the flux gradient with respect to $z$; see ~\ref{seca} for more information on the sign. The forces (and two torques) in the other directions are given by $NI$ times the negative derivative of the flux with respect to the corresponding direction.  Ideally, the geometry of magnet and coil is designed such that the flux through the area of the coil $A_c$ does only change with $z$ and neither with translations along $x$, $y$ nor a small rotation of the coil about any axis. In this case, the force on the coil is purely vertical.

In velocity mode, the electrically open coil moves vertically with the velocity $\ud z/\ud t$.  The induced voltage~$V$ at the coil terminals is measured with a precision voltmeter with high input impedance. It is
\begin{equation}
V = -N \frac{\ud\Phi}{\ud t} = -N \frac{\partial \Phi}{\partial z}\frac{\ud z}{\ud t}.
\label{eq:sec1:V}
\end{equation}
The second identity is only true if the product of the other partial derivatives of $\Phi$  with the corresponding (angular) velocities adds up to zero. But as mentioned above, the magnet-coil system is designed such that the flux derivatives with respect to the coil coordinates, except $z$, are tiny, and, hence, these conditions are met, at least within the uncertainties of the experiment.

Equations~(\ref{eq:sec1:Fz}) and (\ref{eq:sec1:V}) can be combined to what is called the Kibble equation,
\begin{equation}
F_z \frac{\ud z}{\ud t} = V I.
\end{equation}
The Kibble equation connects mechanical power (force times velocity) to electrical power (voltage times current). Electrical power, specifically voltage and resistance, can be measured with quantum electrical standards; see, for example,~\cite{Robinson16,haddad2016bridging}. With these, the link between mass and Planck's constant is complete.

Parts of the literature use the geometric factor $B\ell$, instead of the flux gradient. In this picture, the vectorial version of equation~(\ref{eq:sec1:Fz}) can be written as
\begin{equation}
     \vec{F} = -NI \oint_C \vec{B} \times \ud  \vec{\ell},
\end{equation}
where the force is given by three orthogonal components in the laboratory coordinate system, i.e.,  $\vec{F}=(F_x,F_y,F_z)^T$. The integral is executed over a single wire turn, denoted by $C$ and the line element is given by $\ud  \vec{\ell}$. The direction of the line element is tangential to the turn.  The single  wire turn is considered an average turn in the middle of the coil. For an average coil radius  $r_c$, the length of the integration path  is  $\ell=\oint_C \ud  \ell=2\pi r_c$. Later, surface integrals are introduced. If the surface integral is defined over the coil, it would be calculated over the enclosed area of the single wire turn, $A=\pi r_c^2$. Throughout this text, the wire is assumed to be infinitely thin, i.e., has a negligible cross-section. According to~\cite{Sasso2014}, this approximation  does not introduce significant biases. 

The physics is the same, regardless of whether the derivative of the flux or the geometric factor is used to describe the situation.
Appendix~A in~\cite{Shisong2022} shows how the derivative of a surface integral (flux) can be converted into a line integral with Green's theorem for a divergence-free field. Gauss's law ensures that the divergence of the flux density is zero, i.e., $\vec{\nabla}\cdot\vec{B}=0$. 
Depending on the problem, the geometric factor or the flux derivative may offer an advantage in analyzing it. 
For yoke-based permanent magnets, the flux derivative offers an advantage because the flux through the coil is given by the flux through the inner yoke, and that flux has a simple dependence on $z$, as shown below. In contrast, the reader is encouraged to look at~\cite{li2016coil} which solves some of the integrals in the $B\ell$ picture.
In this article, we will use the flux integral to find the torque on the Kibble coil. As is shown below, the torque can be found by relatively simple geometric considerations and does not require one to solve difficult vector integrals. 

\section{A comparison of two magnet systems}
\label{sec2}

The idea to use the velocity mode to ``calibrate'' the derivative of the flux for the force mode was first published in 1976~\cite{Kibble1976}.
Since then, the Kibble balance, or watt balance as it was previously named, has gone through substantial iterations. 
Many different magnet systems have been used in the last 46 years, e.g. \cite{NPL,nist3,BIPMmag2006,LNEmag,MSL,NISTmag}. A historical overview is given in ~\cite {Shisong2022}. 
In the past two decades, however, a clear favorite has emerged. 
Most groups use a yoke-based permanent magnet system to supply the flux. 

In yoke-based systems, nonlinear interactions between the current in the coil and the iron of the yoke or the permanent magnet material can occur. These interactions are not subject to this article. Here we calculate the force and torques on a current-carrying coil in a given magnetic field without taking  into account the back action of the current on the field or the interaction between the coil and the yoke material. Here, the yoke only shapes the geometry of the field, i.e., the direction and density of the field lines. A summary of the back action can be found in \cite{li2021resolution}. The iron coil interaction is discussed in \cite{li17,NISTmag}.

A yoke-based permanent magnet system can be built in various geometric configurations. Here, we focus on two geometries that we label the BIPM (Bureau International de Poids et Mesures) geometry and the LNE (Laboratoire National de M\'etrologie et d'Essais) geometry. 
The geometries are named after the institutes at which it has been used for the first time in a Kibble balance~\cite{LNEmag,BIPMmag2006}.
The conclusions presented below can easily be transferred by the reader to geometries employed by other Kibble-balance groups~\cite{NPL,NPL3,MSL}.

\begin{figure}[tp!]
\centering
\includegraphics[width=\columnwidth]{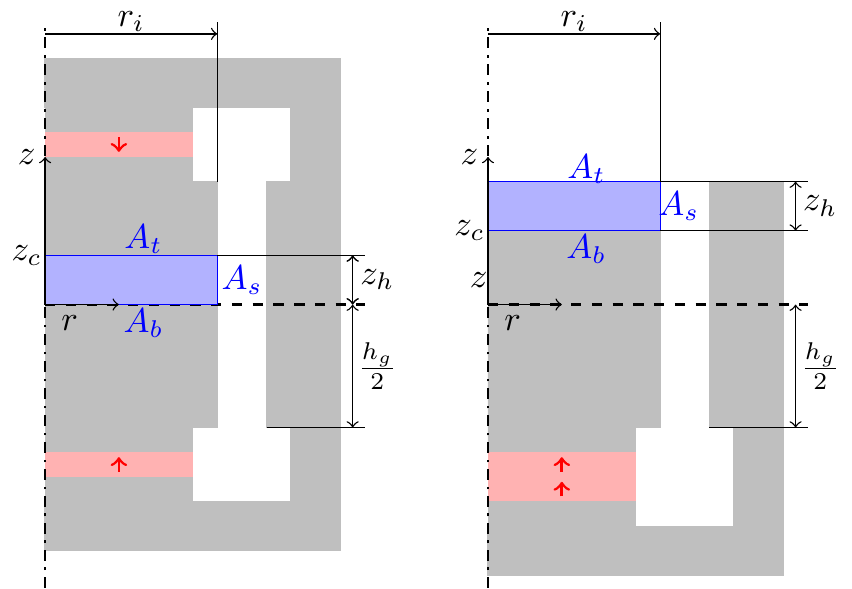}
\caption{Cross-sectional view of the BIPM geometry on the left and the LNE geometry on the right. Both cross-sections are drawn in cylindrical coordinates $r,z$, where $z=0$ is chosen to be in the middle of the air gap with a total height $h_\mathrm{g}$. We assume the air gap dimensions to be the same for both geometries, with $r_\mathrm{i}$ denoting the radius of the inner yoke. The shaded grey areas are the yokes and are made from soft iron. The areas in red represent the permanent magnet material, and the red arrows mark the direction of magnetization. The shaded blue areas (radius $r_\mathrm{i}$, height $z_h$) are cylindrical integration areas for the flux; they are referred to as pillboxes.}
\label{fig:mag:geo}
\end{figure}

A cross-sectional view of the BIPM geometry is shown on the left side of figure~\ref{fig:mag:geo}.
Two permanent magnet disks with opposite magnetization force the magnetic flux radially through the gap.  
The cross-sectional view of the LNE magnet is shown on the right. Here only one permanent magnet disk supplies the flux. 
For this article, the geometries are chosen so that the fluxes in the gaps are approximately the same. 
Hence, the magnet disk in the LNE geometry has approximately the height of both magnet disks in the BIPM magnet.
For simplicity,  fringe fields, which would occur at the end of the air gap, are neglected. This approximation does neither change the  analysis below nor its result\cite{li2021resolution}. 
The air gap dimensions are the same in both cases, with a height of $h_\mathrm{g}$ and an inner radius $r_\mathrm{i}$.
The attentive reader will notice that the LNE magnet slightly differs from the schematic drawing in figure~\ref{fig:mag:geo}. 
In the actual LNE design, the magnetic material is in the outer yoke and not in the inner yoke. 
This detail will not alter the conclusion drawn below. 
Here, we wanted the magnets to be as identical as possible.

We use cylindrical coordinates to describe both magnet systems. 
In both cases, $z=0$ is  in the middle of the gap, and $r=0$ coincides with the symmetry axis of the magnet.
The magnets exhibit perfect azimuthal symmetry, i.e., the magnetic flux density has no dependence on the azimuth, $\phi$.

For both geometries, we presume all the flux from the magnets goes through the air gap (no fringe field as stated above).
We further assume the magnetic flux density in the air gap, near the boundary to the inner yoke $\vec{B}(r_\mathrm{i},z)$ field is horizontal  and constant as a function of $z$.
Note that this can be achieved by engineering the gap width as a function of $z$ as has been done by the researchers at LNE \cite{LNEmag}.

The objective is to obtain the magnetic flux $\Phi_\mathrm{yoke}$ through the cross-sectional area of the inner yoke at position $z_\cc$.  The blue shaded pillboxes drawn in figure~\ref{fig:mag:geo} can be used to calculate the flux. Each pillbox has three surfaces: A circular top $A_t$, a circular bottom $A_b$, with $A_b=A_t$, and a cylindrical surface $A_s$. The vertical separation of top and bottom surface is $z_h$. The top and bottom surfaces have an area of $r_\mathrm{i}^2\pi$ and $A_s=2\pi r_\mathrm{i} z_h$. Since the magnetic flux density is divergence-free, $\vec{\nabla} \cdot \vec{B}=0$, it is
\begin{equation}
    \underbrace{\iint_{A_t} \vec{B} \cdot \ud\vec{A}}_{\Phi_t} +
    \underbrace{\iint_{A_b} \vec{B} \cdot \ud\vec{A}}_{\Phi_b} +
    \underbrace{\iint_{A_s} \vec{B} \cdot \ud\vec{A}}_{\Phi_s} =0
\end{equation}
for all geometries. The convention is that all $\ud\vec{A}$ point outward, i.e., away from the enclosed volume.

For the BIPM geometry, the top surface area is at $z_h=z_\cc$. Hence,
\begin{equation}
\Phi_\mathrm{yoke} (z_h)= \Phi_t = \iint_{A_t} \vec{B} \cdot \ud\vec{A},
\end{equation}
where $\Phi_\mathrm{yoke} (z_h)$ is the magnetic flux through the inner yoke at the plane $z=z_h$.

The bottom surface is at the symmetry plane. Since the magnet is mirror symmetric about $z$, it is $B_z(z)=-B_z(-z)$. Hence, $B_z(0)=0$ and
\begin{equation}
\Phi_b=\iint_{A_b} \vec{B} \cdot \ud\vec{A}=0.
\end{equation}
The flux through $A_s$ is radial because of the high permeability of the yoke material the flux has to be very nearly perpendicular to it. Thus, 
\begin{equation}
\Phi_s(z_h)= \iint_{A_s} \vec{B} \cdot \ud\vec{A}= B_r(r_\mathrm{i}) 2\pi r_\mathrm{i} z_h,
\label{eq:sec2:flux_A_s}
\end{equation}
where $B_r(r_\mathrm{i})$ is the radial magnetic flux at the air/yoke boundary at $r=r_\mathrm{i}$ and $A_s=2\pi r_\mathrm{i} z_h$. Hence, for the BIPM case,
\begin{equation}
\Phi_{\mathrm{yoke,BIPM}} (z_\cc)= -B_r(r_\mathrm{i}) 2\pi r_\mathrm{i} z_\cc,
\label{eq:sec2:flux_yoke_BIPM}
\end{equation}
where we used the fact that the pillbox was chosen with $z_h=z_\cc$.

For the LNE geometry, since per the assumption above, no fringe field exists, the flux through $A_t$ is $\Phi_t=0$.
The flux through the curved surface is the same as for the BIPM magnet, given in equation~(\ref{eq:sec2:flux_A_s}). The bottom surface is traversed by the flux 
\begin{equation}
\Phi_b(z_h) = \iint_{A_b} \vec{B} \cdot \ud\vec{A}= -B_r(r_\mathrm{i}) 2\pi r_\mathrm{i} z_h.
\end{equation}
The bottom surface is at $z_\cc=h_\mathrm{g}/2-z_h$. Thus,
\begin{eqnarray}
    \Phi_{\mathrm{yoke,LNE}} (z_\cc)&=&- \Phi_b(h_\mathrm{g}/2-z_\cc)\nonumber\\
    &=&B_r(r_\mathrm{i}) 2\pi r_\mathrm{i} \left(\frac{h_\mathrm{g}}{2}-z_\cc\right).
    \label{eq:sec2:flux_yoke_LNE}
\end{eqnarray}
The additional negative sign is because the surface normal of $A_b$ of the pillbox points down, but here we calculate the flux in the yoke along the positive $z$ direction.

Inspecting equation~(\ref{eq:sec2:flux_yoke_BIPM}) and 
equation~(\ref{eq:sec2:flux_yoke_LNE}), shows that the negative vertical flux gradients are the same and they agree with the conventional $B\ell$. It is
\begin{equation}
-\frac{\partial \Phi_{\mathrm{yoke,BIPM}}}{\partial z_\cc} 
=
-\frac{\partial \Phi_{\mathrm{yoke,LNE}}}{\partial z_\cc} 
=  B_r(r_\mathrm{i}) 2\pi r_\mathrm{i}. 
\label{eq:sec2:Flux:der}
\end{equation}
We assumed the magnet to produce the same flux density in the gap, and as expected, they would produce the same force on a current-carrying coil.
Note, conventionally the $B\ell$ factor is calculated at the coil radius $B_r(r_\cc)2\pi r_\cc$, but since $B_r \propto 1/r$, it is
\begin{equation}
    B_r(r) 2\pi r = B_r(r_\mathrm{i}) 2\pi r_\mathrm{i} \;\;\mbox{for}\;\; r_\mathrm{i} \le r \le r_\mathrm{i} +d_\mathrm{g}, \label{eq:sec2:Br}
\end{equation}
where $d_\mathrm{g}$ is the width of the gap. Equation~(\ref{eq:sec2:Br}) is valid for a flat field $\partial B_r/\partial z=0$, which is true per our assumption and certainly true at the symmetry plane of the BIPM geometry in an actual magnet.

One notable difference between the two magnet geometries is the flux at $z_\cc=0$. For the BIPM magnet,
\begin{equation}
\Phi_{\mathrm{yoke,BIPM}} (0) =0,
\end{equation}
and for the LNE magnet,
\begin{equation}
\Phi_{\mathrm{yoke,LNE}} (0)= B_r(r_\mathrm{i}) \pi r_\mathrm{i}  h_\mathrm{g}.
\end{equation}

The energy of the coil is given by 
\begin{equation}
E(I)= N I B_r(r_\mathrm{i}) \pi r_\mathrm{i}  h_\mathrm{g}.
\end{equation}
See \ref{seca} for more details on the sign of $E$.
Most notably, the energy is an odd function of $I$. Hence, reversing the current changes the sign of the energy unless the energy was zero before the reversal.  Reversing the current is electrically equivalent to flipping the coil upside down, i.e., rotating the coil by 180$^\circ$ along any horizontal axis. We define this process as \textit{flipping the coil}. When the coil is physically rotated, the electrical terminals of the coil remain connected to the power supply. Hence, if the current was circling clockwise before, it will flow counter-clockwise after the flip and vice versa. In reality, the coil cannot be flipped  because the inner yoke is in the way. However, this obstacle and other practical limitations will not stop us from conducting a thought experiment.

For the LNE magnet, one orientation has significantly lower energy than the other.
 In one orientation, the energy is
\begin{equation}
E(I)= N I B_r(r_\mathrm{i}) \pi r_\mathrm{i}  h_\mathrm{g}.
\end{equation}
In the flipped state, it is
\begin{equation}
E(-I)= -N I B_r(r_\mathrm{i}) \pi r_\mathrm{i}  h_\mathrm{g}.
\end{equation}
This leads to an interesting but fallacious conclusion that by flipping the coil, the energy of the current carrying coil in the LNE magnet changes and, hence, there must be a torque on the slightly tilted coil.

\begin{figure}
\centering
\includegraphics[width=\columnwidth]{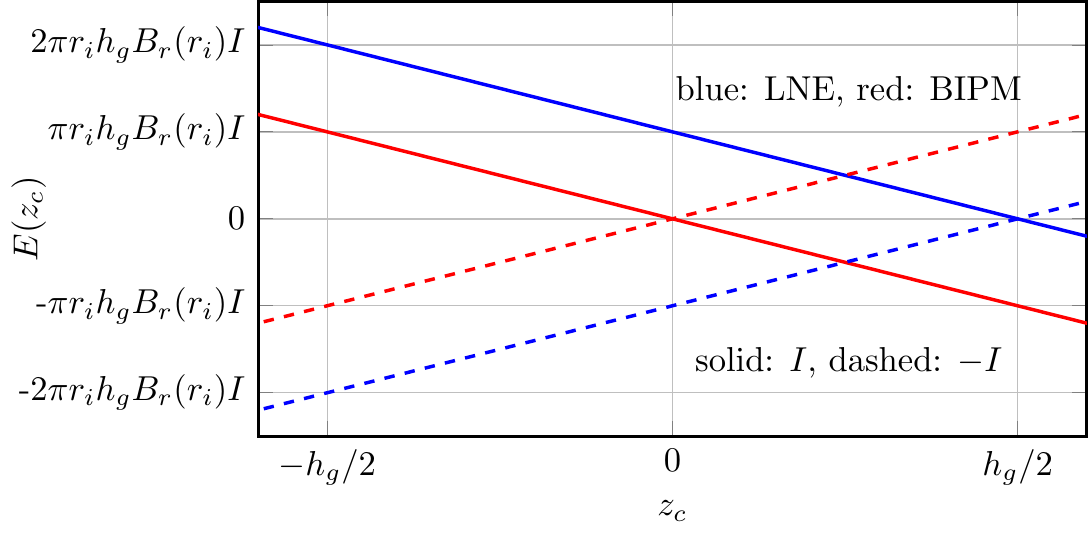}
\caption{The magnetic energy, $E$ as a function of the vertical position of the coil $z_\cc$. The red and blue lines present $E(z_\cc)$ for the BIPM and LNE geometry, respectively. The solid and dashed lines are for different current directions. The two lines with the same current are parallel, i.e., they have the same flux gradient. The topic of discussion is the energy at $z_\cc=0$. The BIPM system has the same energy, $E(0)=0$, independent of the current direction. For the LNE system, the $-I$ configuration has significantly less energy than the $+I$ direction.}
\label{fig:flux:comp}
\end{figure}

The magnetic energy $E$ as a function of $z_\cc$ is shown in figure~\ref{fig:flux:comp}. A coil flip does not change the magnetic flux $\Phi_\mathrm{yoke}$ but changes the energy. The dashed lines in figure~\ref{fig:flux:comp} are mirror images of the solid lines about $E=0$. So, at the weighing position $z_\cc=0$, the magnetic energy of the current-carrying coil in the LNE magnet changes when the current is reversed, but such a change does not occur for the BIPM magnet because it is already $E=0$.

Hence, one would conjecture that if the symmetry is broken (the coil is angled), there would be a  much larger torque on the coil in an LNE magnet  than in the BIPM magnet. That torque would drive the coil to the energetically lower state. This is not the case. The two (idealized) magnets are identical in all aspects. For the LNE magnet, the energy difference corresponds to the mechanical work that would be required to adiabatically move the coil out of the gap and far away, where there is no magnetic flux, and turn it around and bring it back to the middle of the magnet. The torque on an angled coil is the same for the BIPM and the LNE geometry. As the reader may have suspected, only the derivative and not the absolute value of the energy matters, and the derivatives are the same for both magnet systems.

Before we discuss the calculation of the magnetic torque, the direction of the flux inside the inner yoke must be clarified. Since the yoke has a high relative permeability $\mu_r$ (typically ranging from $10^3$ to $10^5$ \cite{NISTmag,LNEmag,hysteresis}), the flux inside the yoke is parallel with its cylinder axis. We assume the cylinder axis of the inner yoke to be aligned with the vertical of the laboratory fixed coordinate system. Because of the high $\mu_r$, the flux density is homogeneous throughout the yoke and  is given by
\begin{equation}
B_{\mathrm{yoke},z} = \frac{\Phi_\mathrm{yoke}}{r_\mathrm{i}^2\pi}.
\end{equation}
For the BIPM magnet, it can be rewritten as
\begin{equation}
B_{\mathrm{yoke},z} = -2 B_r(r_\mathrm{i})  \frac{z_\cc}{r_\mathrm{i}}
\end{equation}

\section{The torques on a tilted coil}
\label{sec3}
The magnetic energy of a current ($I$)  carrying coil with $N$ turns is given by
\begin{equation}
E= -NI \Phi_\mathrm{coil} =- NI \iint_{A_\cc} \vec{B} \cdot \ud\vec{A}.
\label{eq:sec3:E}
\end{equation}
By taking the negative derivative with respect to the angle, the torque on the coil is obtained, 
\begin{equation}
\tau =-\frac{\partial E}{\partial \theta_\cc}= 
 NI \frac{\partial}{\partial \theta_\cc} \iint_{A_\cc} \vec{B} \cdot \ud\vec{A},
\label{eq:sec3:tor}
\end{equation}
where $\theta_\cc$ is the included angle between the $\ud\vec{A}$ and the direction of the yoke, $\vec{e}_z$.

\begin{figure}[tp!]
\centering
\includegraphics[width=0.9\columnwidth]{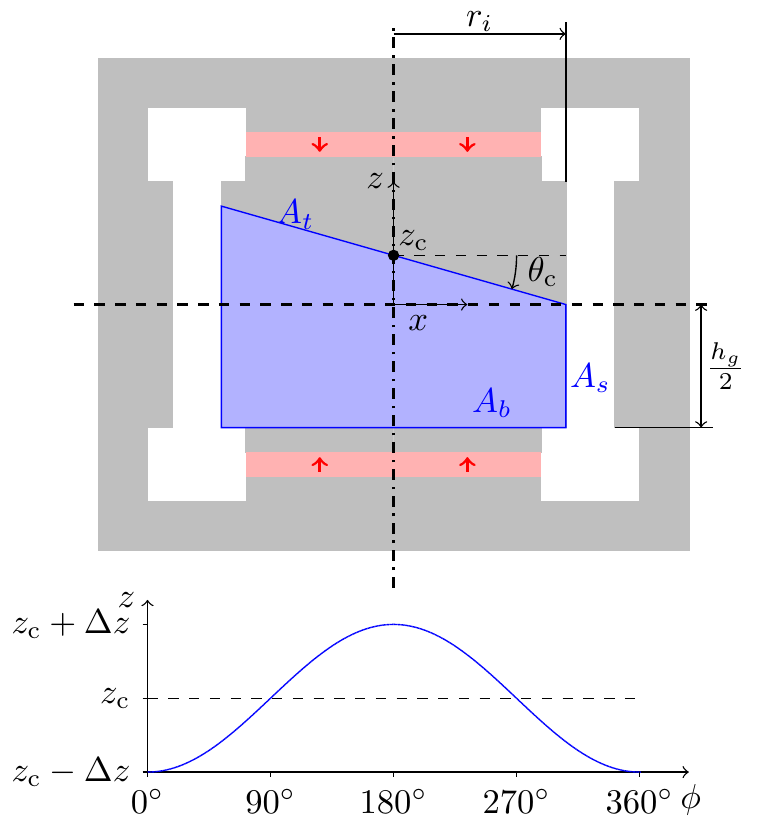}
\caption{The BIPM magnet with a different pillbox. The top surface ($A_t$) of the pillbox is tilted about the $y$-axis by $\theta_\cc$.
The average $z$ position is at $z_\cc$. The lower graph shows the top of $A_s$ as a function of azimuthal angle $\phi$.}
\label{fig:sec3:mag:tilt:pill:box}
\end{figure}

To investigate the change of flux as a function of $\theta_\cc$, a pillbox is used, see figure~\ref{fig:sec3:mag:tilt:pill:box}. The pillbox does not extend to the radius of the coil but only to the radius of the inner yoke. The smaller pillbox simplifies the calculation, and the result is valid for the situation considered here, a flat and purely radial field with the coil and inner yoke being concentric. Later in the text, the pillbox will be expanded to the coil radius.

Now, the top surface is parallel with the coil area, i.e., tilted by $\theta_\cc$ with respect to the horizontal plane, see figure~\ref{fig:sec3:mag:tilt:pill:box}. The upper edge of $A_s$, which coincides with the outer edge of $A_t$ can be written as a function of azimuthal angle $\phi$ and tilt angle $\theta_\cc$. It is
\begin{equation}\label{eq:sec3:height}
    z_t(\phi,\theta_\cc) = z_\cc-r_\mathrm{i}\tan\theta_\cc \cos\phi =z_\cc-\Delta z \cos\phi.
\end{equation}
This equation is plotted in the lower graph of figure~\ref{fig:sec3:mag:tilt:pill:box}.
To shorten the equations, the abbreviation $\Delta z=r_\mathrm{i}\tan\theta_\cc$ is introduced above.
The flux integral through $A_s$ is hence
\begin{equation}
\Phi_s(\theta) =\int_0^{2\pi} \int_{z_b}^{z_t(\phi,\theta_\cc)} B_r(r_\mathrm{i},z) r_\mathrm{i}\, \ud z\,\ud\phi,
\end{equation}
where $z_b$ is the $z$ coordinate of $A_b$.
The $z$ integral can be split in two. One integral from $z_b$ to $z_\cc$ and the other from $z_\cc$ to  $z_t(\phi,\theta_\cc)$. The former integral evaluates the flux through the pillbox with the top at $z=z_\cc$ and $\theta=0$, abbreviated as $\Phi_s(0)$. Hence we obtain
\begin{equation}
\Phi_s(\theta) = \Phi_s(0) +\int_0^{2\pi} \int_{z_\cc}^{z_t(\phi,\theta_\mathrm{c})} B_r(r_\mathrm{i},z) r_\mathrm{i}\, \ud z\,\ud\phi.
\end{equation}

\subsection{Torque on a coil in a magnet with constant field}

First, let's assume $B_r(r_\mathrm{i},z)$ is independent of $z$, hence we replace $B_r(r_\mathrm{i},z)$ with $B_r(r_\mathrm{i},0)$. Then, we can execute the integral over $z$ and we obtain,
\begin{equation}
\Phi_s(\theta) = \Phi_s(0) -\int_0^{2\pi}  B_r(r_\mathrm{i},0) r_\mathrm{i}\Delta z \cos\phi\, \ud\phi.
\end{equation}
The integral over a full period of $\cos\phi$ evaluates to zero. Consequently, the flux through the side surface of the pillbox is independent of a tilt of the top surface. Since the flux through $A_b$ did not change, the flux through $A_t$ must be independent of $\theta_\mathrm{c}$.

If $B_r(r_\mathrm{i},z)$ is independent of $z$, then the flux through the coil surface is independent of $\theta_\cc$, and there is no torque on the current-carrying coil. The statement is also true if the coil is already slightly tilted. We used the BIPM geometry in the figure to show the pillbox, but the result is equally valid for the LNE geometry.

\subsection{Torque on a coil in a magnet with a realistic field profile}

In the section above, we have assumed that the profile is completely flat, i.e. $B_r(r_\mathrm{i},z)=B_r(r_\mathrm{i},0)$ for all $z$. In practice, such a flat field is not achievable. A more realistic assumption is that the profile has a linear and quadratic dependence on $z$ about the symmetry plane which coincides with the nominal coil position in the weighing mode, $z_\cc$. Hence, a Taylor expansion to  second-order around $z_\cc$  yields,
\begin{eqnarray}
    B_r(r_\mathrm{i},z)\approx B_r(r_\mathrm{i},z_\cc) &+& (z-z_\cc)  \left. \frac{\partial B_r}{\partial z}\right|_{z_\cc}\nonumber \\
    &+& \frac{1}{2} (z-z_\cc)^2 \left. \frac{\partial^2 B_r}{\partial z^2}\right|_{z_\cc}.
\end{eqnarray}
Because of $\vec{\nabla}\times \vec{B}=0$, a $z$ dependency of $B_r$ means that the field can no longer be horizontal ($B_z=0$) in the whole gap. Below, we neglect a small but physical necessary $B_z$ term.
The next step is to calculate $\Delta \Phi_s=\Phi_s(\theta) -\Phi_s(0)$. Executing the $z$ integral yields
\begin{eqnarray}
    \int_{z_\cc}^{z_t(\phi,\theta_\mathrm{c})}  B_r(r_\mathrm{i},z)  \ud z =&-& B_r(r_\mathrm{i},0)\Delta z \cos\phi \nonumber\\
    &+& \frac{1}{2} \left.\frac{\partial B_r}{\partial z}\right|_{z_\cc} \Delta z^2 \cos^2\phi\nonumber\\
    &-&\frac{1}{6}\left. \frac{\partial^2 B_r}{\partial z^2}\right|_{z_\cc} \Delta z^3\cos^3\phi.
\end{eqnarray}
Only the middle term remains after integrating over $\phi$ from 0 to $2\pi$. It is,
\begin{equation}
    \Delta \Phi_s = \frac{1}{2} r_\mathrm{i} \pi \Delta z^2 \left. \frac{\partial B_r}{\partial z}  \right|_{z_\cc}.
\end{equation}
The change in flux through $A_t$ is $\Delta \Phi_t=-\Delta \Phi_s$. Hence the torque on the coil can be calculated as
\begin{equation}
\tau = - NI\frac{\partial \Delta \Phi_t}{\partial \theta_\cc}  =-\pi NI r_\mathrm{i}^3  \frac{\sin\theta_\cc}{\cos^3\theta_\cc}\left.\frac{\partial B_r}{\partial z}\right|_{z_\cc}.
\end{equation}
The vertical force on the coil is given by $-NI2\pi r_\mathrm{i} B_r$, so we can use the relative gradient,
\begin{equation}
    g_B:=  \frac{\left.\displaystyle \partial B_r/\partial z\right|_{z_\cc}}{B_r}.
\end{equation}
With this definition and in the small angle approximation ($\sin\theta_\cc\approx\theta_\cc$ and $\cos\theta_\cc\approx1$), the torque on the coil is given by
\begin{equation}
 \tau = \frac{1}{2} F_z r_\mathrm{i}^2 g_B \theta_\mathrm{c}.
 \label{eq:sec3:tor:2}
\end{equation}
Note, for a leveled coil, $\theta_\mathrm{c}=0$,  there is no torque. 

The terms in equation~(\ref{eq:sec3:tor:2}) can be regrouped to help with the dimensional analysis. One $r_\mathrm{i}$ from the square can be lumped with the force, $F_z r_\mathrm{i}$, to give a torque. The other $r_\mathrm{i}$ can be paired with the relative gradient to give a relative change of the profile over the distance $r_\mathrm{i}$, which is, of course, much larger than the typical sweep range. 

The important observation is that if the weighing position is at a maximum or a minimum of $B_r$ with respect to $z$, then there is no torque on the coil no matter how big $\theta_\mathrm{c}$ is. Above, we only used a Taylor expansion up to the second order. The calculation can easily be performed with higher-order terms. Let 
\begin{equation}
 b_k := \left. \frac{\partial^k B_r}{\partial z^k}\right|_{z_\cc},
 \label{eq:sec3:define:bk}
\end{equation}
then the torque on the coil can be written as
\begin{multline}
    \tau = -NI \times \\
    \mathlarger{\sum}_{k,\mathrm{odd}}  r_\mathrm{i}^{k+2} \frac{\ds 2 \pi}{\ds 2^{k+1}} \frac{\ds \left(k+1\right) }{\ds \left(\left(\frac{k+1}{2}\right)!\right)^2} \frac{\ds \tan^{k} \theta_\cc }{\ds \cos^2 \theta_\cc  } b_k.
\end{multline}

Only odd derivatives (odd values of $k$) contribute to the sum. The even derivatives will not produce a flux change and, therefore, will not cause a torque on the coil.
 
\section{Horizontal displacement and tilt of the coil}

\begin{figure}
\centering
\includegraphics[width=0.8\columnwidth]{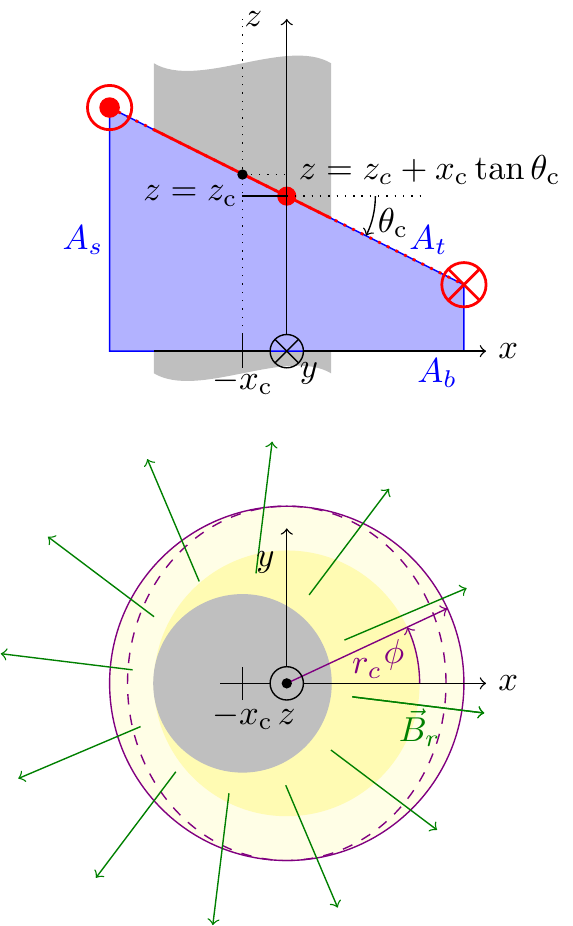}
\caption{ In the upper drawing, the coil (shown in red) is rotated about the $y$ axis by $\theta_\mathrm{c}$ and displaced by $x_\mathrm{c}$ from the symmetry axis of the inner yoke. We establish a pillbox that uses the surface area given by the coil as the top ($A_t$) and a horizontal bottom surface ($A_b$). The curved side wall ($A_s$) connects $A_b$ with $A_t$ to establish a closed pillbox. Unlike before, the coordinate system is centered on the  $A_b$. Hence, the symmetry axis of the yoke goes through $-x_\mathrm{c}$. The lower drawing shows the top view. The coil and the sidewall of the pillbox are on top of each other and are hence drawn in purple. The shape of the top view is an ellipse (dashed line),  with the semi-minor axis being $r_\cc\cos\theta_\cc$. The calculation is carried out along the circular path (solid line). This approximation is good for small angles $\theta_\cc$ to the first order of $\theta_\cc$.  The field lines are centered on the yoke. The yellow shaded areas are explained in the text.}
\label{fig:sec4:tilted:displaced:pillbox}
\end{figure}

As many Kibble balance operators can attest, the most straightforward strategy to minimize a magnetic torque  on a Kibble balance coil is to translate the coil horizontally in the magnet until the torque vanishes.
Changing the magnetic torque on a current-carrying coil by a horizontal displacement relative to the magnet can be intuitively  understood in the $B\ell$ picture. 
The magnetic flux density in the air gap falls like $B(r)\propto 1/r$. 
Hence by displacing the coil, some arc sections of the coil experience a weaker while other sections a stronger force, and a torque is generated. Visibly, this effect is indicated in the lower drawing of figure~\ref{fig:sec4:tilted:displaced:pillbox}. The density of the green arrows crossing the blue circles is higher on the left than on the right side of the coil center. So more force per unit wire length is generated  on the left, and a torque arises that would rotate the coil about the $y$ axis.

How can this fact be explained with the flux gradient picture? Naively, one would think that the flux does not change if the coil is translated. It seems that the flux through the coil is given by the flux through the inner yoke. And since the coil opening always includes the inner yoke, the flux would not change. This line of thinking is wrong, as shown below.

The origin of the torque can be understood with  another pillbox, shown in figure~\ref{fig:sec4:tilted:displaced:pillbox}. This time the pillbox has a larger radius. The radius is that of the coil. The top surface $A_t$ is comprised of the coil. The bottom surface $A_b$ is a projection down of the top surface to $z=0$ or another reference plane. The curved side surface connects $A_b$ with $A_t$ and completes the pillbox.
Since the total flux is 0, i.e.,
\begin{equation}
\Phi_\mathrm{t}+\Phi_\mathrm{s}+\underbrace{\Phi_\mathrm{b}}_{\mathrm{const.}} =0 \;\;\mbox{hence}\;\; \Phi_\mathrm{t} = -\Phi_\mathrm{s} - \Phi_\mathrm{b}.
    \label{eq:sec4:fluxsum}
\end{equation}
As indicated, the flux through the bottom is constant, and its value is not important.
The aim of the following few paragraphs is to calculate $\Phi_s$. We choose a coordinate system that is different from the ones used in the previous section. The origin of the coordinate system is in the middle of $A_b$, which makes it simple to parameterize the pillbox. In this coordinate system, the symmetry axis of the inner yoke intersects the $xy$-plane at $(-x_\cc,0)$.

The height of the cylindrical sidewall is $z_\mathrm{t}(\phi)$, as given  by equation~(\ref{eq:sec3:height}) where $r_\mathrm{i}$ is replaced by $r_\cc$.
In the chosen coordinate system, the sidewall of the pillbox is given by
\begin{equation}
    \vec{r}=r_\cc \vec{e_r} =r_\cc
    \left(
    \begin{array}{l}
    \cos\phi \\
    \sin\phi\\
    0
    \end{array}
    \right).
\end{equation}
The pillbox is parameterized with a circular cross-section, which is an approximation. In reality, the projection of the tilted coil onto the $xy$-plane results in an ellipse with a major semi-axis along $y$ of length $r_\cc$ and a minor semi-axis along $x$ of length $r_\cc \cos\theta_\cc$. For small angles, $\theta_\cc$ the minor axis approximates $r_\cc(1-\theta_\cc^2/2)$ and is to first-order independent of $\theta_\cc$. Hence, using a circular cross-section will be correct up to the first order.

The flux penetrating the cylindrical wall is 
\begin{multline}
\vec{B}(r)\ud \phi =\\ \frac{B_r(r_\mathrm{i}) r_\mathrm{i}} {(x_\cc+r_\cc \cos\phi)^2+r_\cc^2 \sin^2\phi}
    \left(
    \begin{array}{l}
    x_\cc+r_\cc \cos\phi \\
    r_\cc \sin\phi\\
    0
    \end{array}
    \right).
    \label{eq:sec4:Br}
\end{multline}
Here, the flux density  $B_r(r_\mathrm{i})$ is assumed to be independent of $z$.
Then, the total flux through the sidewall is
\begin{equation}
\Phi_s =\int_0^{2\pi} \vec{B} \cdot \vec{r}\, z_\mathrm{t}(\phi)\,  \ud \phi,
\label{eq:sec4:integral}
\end{equation}
where the term in the integral is given by
\begin{multline}
\vec{B} \cdot \vec{r}\, z_\mathrm{t}(\phi)\  = \\
B_r(r_\mathrm{i}) r_\mathrm{i} r_\cc \frac{\left(r_\cc+x_\cc \cos\phi\right)\left(z_\cc-r_\cc \cos\phi \tan\theta_\cc\right)}{r_\cc^2+x_\cc^2+2r_\cc x_\cc\cos\phi}.
\label{eq:sec4:integral:result}
\end{multline}
Due to the  symmetry of the problem, it is sufficient to execute the integral for $\phi$ from 0 to $\pi$ and double the result. For $x_\cc<r_\cc$ it is,
\begin{equation}
\Phi_\mathrm{s} =B_r \pi r_\mathrm{i} (2 z_\cc + x_\cc \tan\theta_\cc).
\label{eq:sec4:flux:top}
\end{equation}

Then, according to equation~(\ref{eq:sec4:fluxsum}) the flux through the top of the pillbox is
\begin{equation}
\Phi_\mathrm{t} =-B_r 2 \pi r_\mathrm{i} \left( z_\cc + \frac{x_\cc \tan\theta_\cc}{2}\right) -\Phi_\mathrm{b} .
\end{equation}
Since, $F_z = -NI B_r 2\pi r_\mathrm{i}$ we can replace $-B_r 2 \pi r_\mathrm{i}$ with $F_z/NI$.   Since the energy is $E=-NI\Phi_t$ we have
\begin{equation}
E =-F_z  \left(z_\cc + \frac{x_\cc \tan\theta_\cc}{2}\right)+NI\Phi_\mathrm{b},
\end{equation}
where $NI\Phi_\mathrm{b}$ is a constant that is irrelevant.
Using the small angle approximation, $\tan\theta_\cc\approx\theta_\cc$, the horizontal force and torque about the $y$ direction can be calculated as
\begin{equation}
F_x=-\frac{\partial E}{\partial x_\cc} = \frac{F_z}{2} \theta_\cc
\label{eq:sec4:Fx:1}
\end{equation}
and
\begin{equation}
\tau_\theta=-\frac{\partial E}{\partial \theta_\cc} = \frac{F_z}{2} x_\cc.
\label{eq:sec4:tau:1}
\end{equation}

Interestingly, if one were to calculate the flux that goes through the intersection of the coil and the yoke, one would obtain
\begin{equation}
\Phi_\mathrm{top,yoke} =-B_r 2 \pi r_\mathrm{i} ( z_\cc + x_\cc \tan\theta_\cc),
\label{eq:sec4:yoke_wrong}
\end{equation}
which is equation~(\ref{eq:sec2:flux_yoke_BIPM}) evaluated at $z_\cc + x_\cc \tan\theta_\cc$. 
The $z_\cc$ value in equation~(\ref{eq:sec2:flux_yoke_BIPM})  must be modified because the coil plane meets the symmetry axis of the yoke by $x_\cc \tan\theta_\cc$  above $z_\cc$ as is indicated by the black dot in figure~\ref{fig:sec4:tilted:displaced:pillbox}.
If one were to use equation~(\ref{eq:sec4:yoke_wrong}) instead of 
equation~(\ref{eq:sec4:flux:top}) to calculate $F_x$ and $\tau_\theta$, each result would be double.

The difference in flux is
\begin{equation}
\Phi_\mathrm{leakage}= \Phi_\mathrm{t}-
\Phi_\mathrm{top,yoke}  = B_r 2 \pi r_\mathrm{i} \frac{x_\cc \tan\theta_\cc}{2}.
\label{eq:sec4:flux:leakage}
\end{equation}
for $\Phi_b=0$, which is the case if the bottom of the pillbox is chosen at the symmetry plane of the BIPM magnet.

This missing flux must escape through the area enclosed by the coil that is not filled with the inner yoke indicated by the yellow (both shades) area  in figure~\ref{fig:sec4:tilted:displaced:pillbox}. Interestingly, $\Phi_\mathrm{leakage}$ does not depend on $r_\cc$, but only $r_\mathrm{i}$. Hence, one can hypothetically shrink the coil and have the same flux leakage. Therefore the flux must escape in the area that is shaded in dark yellow in figure~\ref{fig:sec4:tilted:displaced:pillbox}. It is the smallest circle that is concentric with the coil but fully contains the yoke.

The flux leakage only occurs because the coil is not centered on the yoke. For a centered coil, no flux escapes through the air part of the coil surface because the loss of flux on one side of the tilt axis is compensated by the gain of flux on the other side. For this reason, the pillbox was correctly constructed about the inner yoke in previous sections. 

Based on equations~(\ref{eq:sec4:Fx:1}) and (\ref{eq:sec4:tau:1}),
the translation and tilt are coupled equations. A translated current-carrying coil will produce a torque that produces a tilt. Likewise, a coil that is tilted and carries current will produce  a force that will further translate the coil causing tilt. This result in this section is only valid in the $xz$-plane. The complete two-dimensional case is solved in \ref{secb}. 

The difference between the torques discussed in section~\ref{sec3} and here is that  the torques discussed in section~\ref{sec3} arise from a finite first derivative in the profile, whereas the torques discussed in this section result from the coil not being centered in the magnet. In practice, both torques need to be added together to understand the situation in the laboratory.

\section{Motion of the coil in the magnet}

One result obtained above is that the torque and horizontal forces on a coil are coupled. What is the final position of the current carrying coil? We assume that without current the coil coordinates are given by $\vec{p_o} =(x_o,\theta_o)^T$. The goal is to find the coordinates $\vec{p_\cc} =(x_\cc,\theta_\cc)^T$ of the current carrying coil. The current in the coil is such that it produces the vertical force $F_z$.   We assume that the stiffness of the suspension is $k_x$ for translation and $\kappa_\theta$ for rotation.
 The mechanical and magnetic energy in the limit of small $\theta_\cc$  of the system is given by
\begin{equation}
E = -\frac{1}{2} F_z x_\mathrm{c} \theta_\mathrm{c} + \frac{1}{2} k_x (x_\mathrm{c}-x_o)^2 + \frac{1}{2} \kappa_\theta (\theta_\mathrm{c}-\theta_o)^2.
\label{eq:sec5:energy}
\end{equation}
The origin of the coordinate system  ($x=0,\theta=0$) is chosen based on the symmetry of the magnet system, as it has been in the previous sections.

With current in the coil,
the  equilibrium position is where the partial derivatives of the energy with respect to $x_\cc$ and $\theta_\cc$ are zero. It is
\begin{eqnarray}
\frac{\partial E}{\partial x_\mathrm{c}}&=& 
-\frac{1}{2} F_z \theta_\mathrm{c} +  k_x x_\mathrm{c}-k_x x_o =0 \;\mbox{and} \\
\frac{\partial E}{\partial \theta_\mathrm{c}}&=& 
-\frac{1}{2}F_z x_\mathrm{c} +  \kappa_\theta \theta_\mathrm{c}-\kappa_\theta \theta_o  =0. 
\label{sec5:deriv:eq:zero}
\end{eqnarray}
Succinctly in a matrix equation, it is
\begin{equation}
\left(\begin{array}{cc} k_x & -\frac{1}{2} F_z \\ -\frac{1}{2} F_z &\kappa_\theta\end{array}\right)
\left(\begin{array}{cc} x_\mathrm{c} \\ \theta_\mathrm{c}\end{array}\right) =
\left(\begin{array}{cc} k_x x_o \\ \kappa_\theta \theta_o\end{array}\right).
\label{sec5:matrix}
\end{equation}
Multiplying both sides with the inverted matrix yields
\begin{equation}
\left(\begin{array}{cc} x_\mathrm{c} \\ \theta_\mathrm{c}\end{array}\right) =
\frac{2}{4k_x\kappa_\theta -F_z^2}
\left(\begin{array}{l} 2k_x \kappa_\theta x_o +  \kappa_\theta \theta_o F_z \\ 2k_x \kappa_\theta \theta_o+k_x x_o  F_z  \end{array}\right).
\label{eq:sec5:matrix:sol}
\end{equation}
Introducing the displacement matrix,
\begin{equation}
\mathbf{D} =
\frac{2}{4k_x\kappa_\theta -F_z^2}
\left(\begin{array}{ll} 2k_x \kappa_\theta &  \kappa_\theta  F_z \\ 
k_x  F_z & 2k_x \kappa_\theta   \end{array}\right)
\label{eq:sec5:displacement:matrix}
\end{equation}
simplifies equation~(\ref{eq:sec5:matrix:sol}) to
\begin{equation}
\vec{p}_\cc = \mathbf{D}\, \vec{p}_o.
\end{equation}
Different scenarios of its solution  will be discussed in the next section.
\subsection{Different suspensions}
The final equilibrium position of the current-carrying coil depends on the denominator in equation~(\ref{eq:sec5:displacement:matrix}). Three cases can be distinguished: The unstable case, i.e.,  $F_z^2= 4k_x \kappa_\theta$, the soft suspension $4k_x\kappa_\theta<<F_z^2$, and the stiff suspension $4k_x\kappa_\theta>>F_z^2$. Each case is discussed below. Note, that for all suspensions, the coil is connected to the Kibble balance in a compliant fashion. This is different from the idea put forward by Kibble and Robinson~\cite{Kibble_2014}, where the coil motion is completely constrained to a one-dimensional motion.

\subsection{The unstable case}
The theoretical solutions for this case would be $x_\cc=\infty$ and $\theta_\cc=\infty$. However, in reality, the nonlinear behavior of the suspension will alter the product $k_x\kappa$ as $x_\cc$ and $\theta_\cc$ increase such that a stable equilibrium will occur. Besides stability, the fine-tuning required  to obtain $F_z^2= 4k_x \kappa_\theta$ would be delicate and, hence, never occur in practice. 

But also a small value,  $|4k_x \kappa_\theta/F_z^2-1|$ is bad, as it will lead to large values of $|x_\cc|$ and $|\theta_\cc|$. The sign of these variables depends on the sign of $4k_x\kappa_\theta -F_z^2$. 

\subsection{The stiff suspension}
In contrast to the case discussed above, the solution to this case is intuitive and stable. Since $4k_x\kappa_\theta>>F_z^2$ we can neglect $F_z$ in the denominator of equation~(\ref{eq:sec5:displacement:matrix}) and obtain
\begin{equation}
\mathbf{D} =
\left(\begin{array}{ll} 
1 & F_z/(2k_x)\\
F_z/(2\kappa_\theta )& 1 \\ 
 \end{array}\right).
 \label{eq:sec5:matrix:sol:stiff}
\end{equation}

The final position of the current-carrying coil deviates from the position of the coil without current by terms that are proportional to $F_z$ and the misalignment in the other channel ($x_o$ for $\theta_\cc$ and $\theta_o$ for $x_\cc$).  The change in each coordinate is simply given by $F_z/2$ divided by the respective spring constant times the position without current in the other coordinate, i.e.,  $x_\cc-x_o= 
 \theta_o F_z/(2k_x) $ and $\theta_\cc-\theta_o = x_o F_z/(2\kappa)$.
Changing the direction of $F_z$ would simply change the direction of the deviation, e.g., $x_\cc-x_o$. Note, $F_z=0$ is included in this case.

\subsection{The soft suspension}
The soft suspension  ($4k_x\kappa_\theta<<F_z^2$) will also give a stable solution. This solution, however, is less intuitive. Here, the term $4k_x\kappa_\theta$ in the denominator can be ignored, and~(\ref{eq:sec5:displacement:matrix}) is,
\begin{equation}
\mathbf{D} =
\frac{-2}{ F_z^2}
\left(\begin{array}{ll} 2k_x \kappa_\theta &  \kappa_\theta  F_z \\ 
k_x  F_z & 2k_x \kappa_\theta   \end{array}\right).
\label{eq:sec5:matrix:sol:soft}
\end{equation}

In the limit of very large $|F_z|$ both components of $\vec{p}_\cc$ converge to 0. The large magnetic force in the $\pm z$ direction straightens out the flexure joints.

For smaller $|F_z|$, it is best to discuss each parameter by setting the other initial parameter to zero. For $x_o=0$ it is 
\begin{equation}
\vec{p}_\cc =
-2\kappa_\theta \theta_o
\left(\begin{array}{l} 
1/ F_z \\
2k_x /F_z^2 \end{array}\right),
\label{eq:sec5:matrix:sol:stiff:x0}
\end{equation}
and for $\theta_o=0$ it is
\begin{equation}
\vec{p}_\cc =
-2k_x x_o
\left(\begin{array}{l} 
2 \kappa_\theta/ F_z^2 \\
1 /F_z \end{array}\right).
\label{eq:sec5:matrix:sol:stiff:theta0}
\end{equation}
Adding equation~(\ref{eq:sec5:matrix:sol:stiff:x0}) and equation~(\ref{eq:sec5:matrix:sol:stiff:theta0}) returns the general solution, i.e., 
$\mathbf{D} $ from equation~(\ref{eq:sec5:matrix:sol:soft}) multiplied by the equilibrium position $\vec{p_o} =(x_o,\theta_o)^T$.
We see that the aligned coil coordinate moves with a term that is proportional to $-1/F_z$, whereas the other one is proportional to $1/F_z^2$. Interestingly that means, if one compares the position of the coil for mass on, $F_z=mg/2$ with  the position for mass off $F_z=-mg/2$, one would not see a difference for the coordinate that is proportional to $1/F_z^2$. So, there could be a horizontal force or torque on the current-carrying coil and it is not detected because it has the same sign and magnitude for opposite current directions.

\subsection{Discussion of the three scenarios}
\begin{figure}
\centering
\includegraphics[width=\columnwidth]{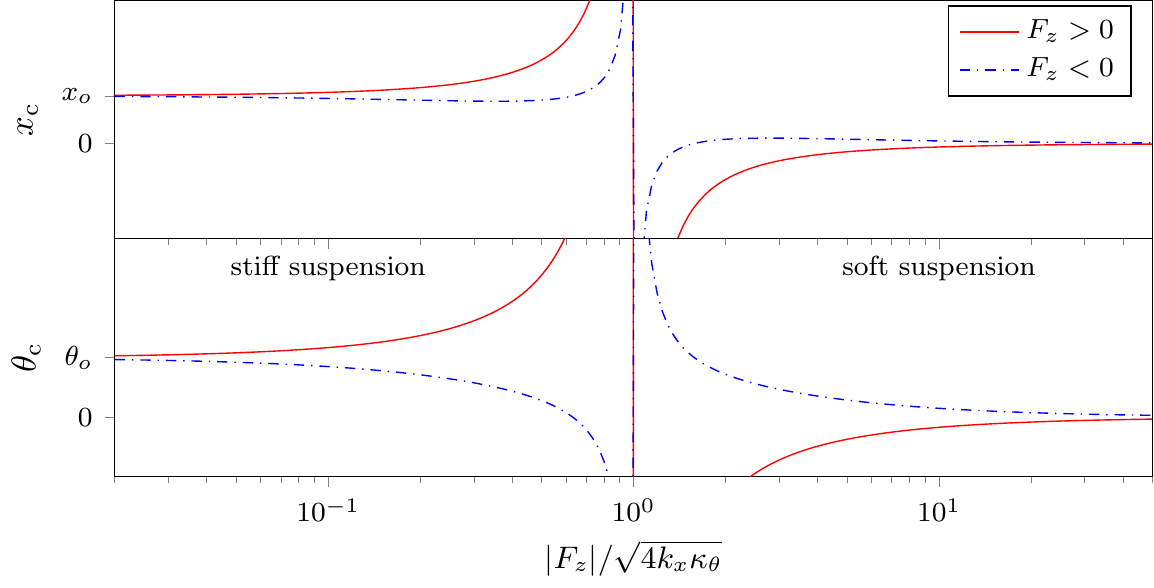}
\caption{The equilibrium position for the current carrying coil as a function of $F_z/\sqrt{4\kappa_\theta k_x}$. 
Both $x_o$ and $\theta_o$ have small positive values. Each value is indicated by the tick mark in the respective vertical axis.}
\label{fig:sec5:theo:coil:pos}
\end{figure}	

Figure~\ref{fig:sec5:theo:coil:pos} shows the equilibrium position of the coil as a function of 
$F_z/\sqrt{4\kappa_\theta k_x}$ for positive and negative $F_z$. A few characteristics are clearly visible in the graphs in the figure. The equilibrium position of the current carrying coil for small $|F_z|$ is $\vec{p}_o$ and for large $|F_z|$ it is $(0,0)^T$. There is a pole at $F_z=\sqrt{4\kappa_\theta k_x}$. At the pole, the order of the variables changes for the two current directions. For the chosen initial conditions and the $x_\cc$ coordinate, the trace with $F_z>0$ is above the one for $F_z<0$ on the left side of the pole and below on the right side of the pole.

\section{Obtaining coil misalignment by experiment}

With the equations discussed in the previous section, the coordinates of the energized coil can be calculated from the coordinates ($\vec{p}_o$) of the de-energized coil. These equations, however, require the absolute knowledge of $\vec{p}_o$ with respect to the coordinate system that is centered on the magnet. Most Kibble balances employ optical levers and $xy$ detectors to monitor the angle and position of the coil. The outputs of these sensors are relative to an arbitrary null of the photodetector and do not provide absolute position and orientation of the electrical center of the coil with respect to the magnetic center of the magnet. 

Here, we show that it is possible to gain that information with three measurements that are already being made in a Kibble balance experiment. In the velocity mode, the coil has no current, so the measurement of the coil position, $\vec{m}_o$ with $F_z=0$ is made. The other two measurements are obtained in force mode which is usually comprised of mass-on and mass-off measurements. In the mass-on measurement, the coil provides $F_z=mg/2$ and we denote the measured coil position by $\vec{m}_+$. Similarly, $F_z=-mg/2$ for the mass-off measurement. In this case, the measured coil position is  abbreviated as $\vec{m}_-$.

Equation~(\ref{eq:sec5:matrix:sol}) can be rewritten by introducing the unitless parameter $f := F_z/\sqrt{4\kappa_\theta k_x}$. The  result depends on the square root of the product and quotient of $k_x$ and $\kappa_\theta$. So, two more variables can be introduced, $\xi=\sqrt{\kappa_\theta/k_x}$ and $\eta=\sqrt{k_x\kappa_\theta}$. The dimensions of $\xi$ and $\eta$ are length and force, respectively. With the new variables, equation~(\ref{eq:sec5:matrix:sol}) becomes
\begin{equation}
\left(\begin{array}{cc} x_\mathrm{c} \\ \theta_\mathrm{c}\end{array}\right) =
\frac{1}{\ds 1 -f^2}
\left(\begin{array}{l}
 x_o + \theta_o f\xi  \\
\theta_o+x_o f/\xi 
\end{array}\right).
\label{eq:sec6:matrix:sol3}
\end{equation}
The measured values must include the unknown offsets in both rows. So, the measurements are
\begin{equation}
\vec{m}_\mathrm{state}=
\frac{1}{\ds 1 -f^2}
\left(\begin{array}{l}
 x_o + \theta_o f\xi  \\
\theta_o+x_o f/\xi 
\end{array}\right) +
\left(\begin{array}{cc} x_\mathrm{off} \\ \theta_\mathrm{off}\end{array}\right).
\label{eq:sec6:matrix:sol:meas}
\end{equation} 
The three types of measurement are distinguished by the subscript $_\mathrm{state}$. The state is either $+$, $o$, or $-$ depending on whether $F_z$  is $+mg/2$, $0$, or $-mg/2$. For these three cases, the unitless parameter is $+f$, 0, or $-f$, respectively.
From these three measurements, two differences can be calculated.
\begin{equation}
\vec{\Delta}_{+-}= \vec{m}_+-\vec{m}_-
=
\frac{2f}{\ds 1 -f^2}
\left(\begin{array}{l}
 \theta_o\xi  \\
 x_o /\xi 
\end{array}\right).
\label{eq:sec6:delta_pm}
\end{equation}
and
\begin{equation}
\vec{\Delta}_{+0}= \vec{m}_+-\vec{m}_o=
\frac{f}{\ds 1 -f^2}
\left(\begin{array}{l}
 x_o f+\theta_o\xi  \\
 \theta_o f +x_o/ \xi 
\end{array}\right).
\label{eq:sec6:delta_p0}
\end{equation}
Next, a linear combination of the two differences is used,
\begin{equation}
\vec{\Delta}_\mathrm{aux}:=
\vec{\Delta}_{+0}-\frac{\ds \vec{\Delta}_{+-}}{\ds 2}=
\frac{f^2}{\ds 1 -f^2}
\left(\begin{array}{l}
 x_o  \\
 \theta_o  
\end{array}\right).
\label{eq:sec6:delta_aux}
\end{equation}

With these results, $f$ and $\xi$ can be obtained. They are
\begin{equation}
f = 2\sqrt{\bigg|\frac{\vec{\Delta}_\mathrm{aux}[1]}{\vec{\Delta}_{+-}[1]} 
\frac{\vec{\Delta}_\mathrm{aux}[2]}{\vec{\Delta}_{+-}[2]}\bigg| }
\label{eq:sec6:f}
\end{equation}
and
\begin{equation}
\xi= \sqrt{\bigg|\frac{\vec{\Delta}_\mathrm{aux}[1]}{\vec{\Delta}_\mathrm{aux}[2]} 
\frac{\vec{\Delta}_{+-}[1]}{\vec{\Delta}_{+-}[2]}\bigg|}.
\label{eq:sec6:xi}
\end{equation}
The notation $\Delta_x[1]$ and $\Delta_x[2]$ is used to indicate the first and second row of $\Delta_x$, with $_x$ being either $_{+-}$ or $_\mathrm{aux}$.
 Once $\xi$ and $f$ are obtained, equation~(\ref{eq:sec6:delta_pm}) can be solved for $x_o$ and $\theta_o$. It is
\begin{equation}
\left(\begin{array}{l}
 x_o \\
 \theta_o  
\end{array}\right)
=
\frac{\ds 1 -f^2}{2f}
\left(\begin{array}{l}
 \vec{\Delta}_{+-}[2]  \xi\\
 \vec{\Delta}_{+-}[1] /\xi
 \end{array}\right).
\label{eq:sec6:xo:thetao:solution}
\end{equation}
Since $F_z$ is known, a value for $\eta$ can be obtained via $\eta=F_z/(2f)$. Finally, from $\eta$ and $\xi$ both stiffnesses can be calculated. They are
\begin{equation}
k_x  = \eta\, /\xi \;\mbox{and}\; \kappa_\theta = \eta\xi
\label{eq:sec6:kx:kappa}
\end{equation}

\begin{figure}
\centering
\includegraphics[width=\columnwidth]{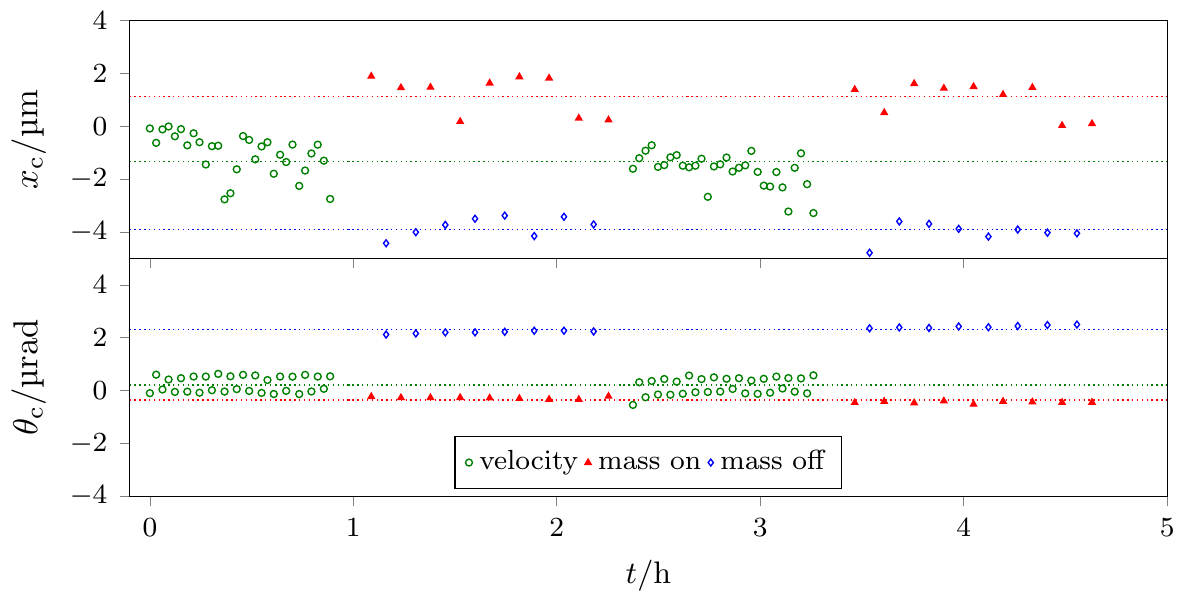}
\caption{
For the data shown here, the alignment of the Kibble balance was not ideal. Position measurements for a typical alignment can be found in figure~9 of \cite{haddad2016invited}.}
\label{fig:sec6:coil:pos:NIST4}
\end{figure}

Figure~\ref{fig:sec6:coil:pos:NIST4} shows a typical measurement of these positions as they were carried out with the Kibble balance at NIST, named NIST-4~\cite{haddad2016invited}. For the data shown here, the Kibble balance was not particularly well aligned. The top graph shows the measured coil position for the three measurement types, velocity mode (in green), mass-on (in red), and mass-off (in blue). The dotted horizontal lines show the average values for the first 5 hours of the run. There is noise on the detectors, but the different coil positions for the three measurements can be distinguished. From these averaged values, $x_o$ and $\theta_o$ can be calculated. The result and the results of the intermediate steps are shown in table~\ref{tab:sec6:results}.

\begin{table}[h!]
\begin{center}
\caption{Calculation of $x_o$ and $\theta_o$ from the measured values shown in figure~\ref{fig:sec6:coil:pos:NIST4}. The top table shows the measured position vectors for mass-on, mass-off, and velocity mode and the differences that were obtained from these measurements. The last line shows the result. The lower table shows the values that were obtained for the ancillary variables. The digits in parentheses give the one standard deviation uncertainty in the last or last two digits of the corresponding value.}

\begin{tabular}{ l S[table-format=4.2(2)] S[table-format=4.2(2)]  l}
\hline
& \multicolumn{1}{c}{$x/\SI{}{\micro\meter}$} & \multicolumn{1}{c}{$\theta/\SI{}{\micro\radian} $} & source\\
\hline
$\vec{m}_+$ &  1.12 (16) & -0.36(2)&measured\\
$\vec{m}_o$ &  -1.34(10)&  0.21(4)&measured\\
$\vec{m}_-$ & -3.90(10) &  2.32(3)&measured\\
$\vec{\Delta}_{+-}$& 5.02(19)  & -2.67(4) &Eq.~(\ref{eq:sec6:delta_pm})\\
$\vec{\Delta}_{+0}$&2.46(19)   &-0.57(4)&Eq.~(\ref{eq:sec6:delta_p0})\\
$\vec{\Delta}_\mathrm{aux}$&-0.05(14) & 0.77(4)&Eq.~(\ref{eq:sec6:delta_aux})\\
$x_o$,$\theta_o$&  -4.31(33) &65(49) &Eq.~(\ref{eq:sec6:xo:thetao:solution})\\
\hline
quantity & \multicolumn{2}{c}{value} & source\\
\hline
$f$ & 0.11(7) & &Eq.~(\ref{eq:sec6:f})\\
$\xi$&0.35(22) &\multicolumn{1}{l}{\si{\meter}}&Eq.~(\ref{eq:sec6:xi})\\
$k_x$ &64(18) &\multicolumn{1}{l}{\si{\newton \per \meter}} &Eq.~(\ref{eq:sec6:kx:kappa})\\
$\kappa_\theta$ &8.0(6) &\multicolumn{1}{l}{\si{\newton \meter \per \radian}} &Eq.~(\ref{eq:sec6:kx:kappa})\\
$\sqrt{4  k_x \kappa_\theta} $ & 45(19)&\multicolumn{1}{l}{\si{\newton }}&last 2 lines\\
$F_z=mg/2$ & 4.904 &\multicolumn{1}{l}{\si{\newton }}& measured\\
\hline
\end{tabular}
\label{tab:sec6:results}
\end{center}
\end{table}

\begin{figure}[!ht]
\centering
\includegraphics[width=\columnwidth]{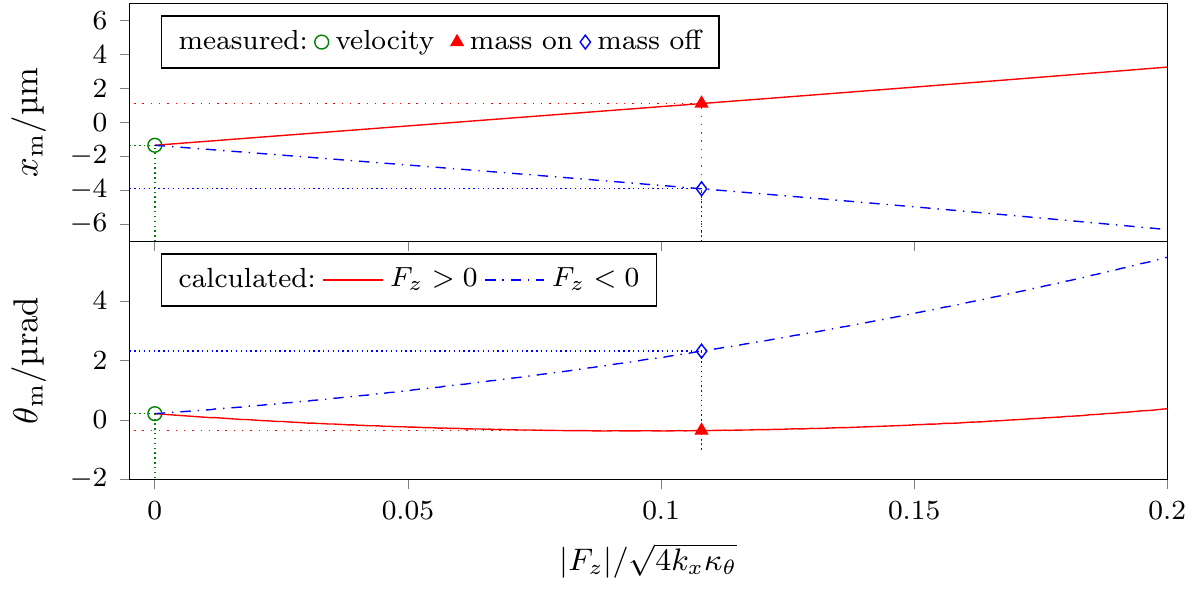}
\caption{The measured coil position of the coil in the NIST-4 Kibble balance as a function of $F_z/\sqrt{4k_x\kappa_\theta}$. This calculation is performed with the parameters given in table~\ref{tab:sec6:results}. For the measurements shown here the Kibble balance measured a  1\,kg mass and hence $f=0.108$. The measured coil positions are indicated by markers. At $f=0$, the reported positions are not 0, caused by offsets in the position measurement system.}
\label{fig:sec6:eta:nist4}
\end{figure}	

From the results, it can be concluded that the coil center is off by $\SI{4.3}{\micro\meter}\, \pm\, \SI{0.3}{\micro\meter}$ from the center of the magnet. Furthermore, the coil is titled by $\SI{65}{\micro \radian}\, \pm\, \SI{49}{\micro \radian}$ from the horizontal reference plane given by the magnet. Having these numbers makes it straightforward to align the Kibble balance.

Figure~\ref{fig:sec6:eta:nist4} shows similar content as figure~\ref{fig:sec5:theo:coil:pos}. But this time the calculation is performed with the values reported in table~\ref{tab:sec6:results}. It can be seen that NIST-4 operates in the regime of the stiff suspension $|F_z| \approx 0.1 \sqrt{4 k_x \kappa_\theta}$ if it is used to weigh a \SI{1}{\kilo \gram} mass. The measurements of the position detectors are also indicated in the figure.

\subsection{On the stiffness of the suspension}
Besides  $x_o$ and $\theta_o$ the procedure above yields two stiffnesses of the coil suspension, $k_x=\SI{64}{\newton \per \meter}$\,$\,\pm\, \SI{18}{\newton \per \meter}$ and $\kappa_\theta=\SI{8.0}{\newton \meter \per \radian}\,\pm\,$ $\SI{0.6}{\newton \meter \per \radian}$.  How reasonable are these values?

\begin{figure}
\centering
\includegraphics[width=\columnwidth]{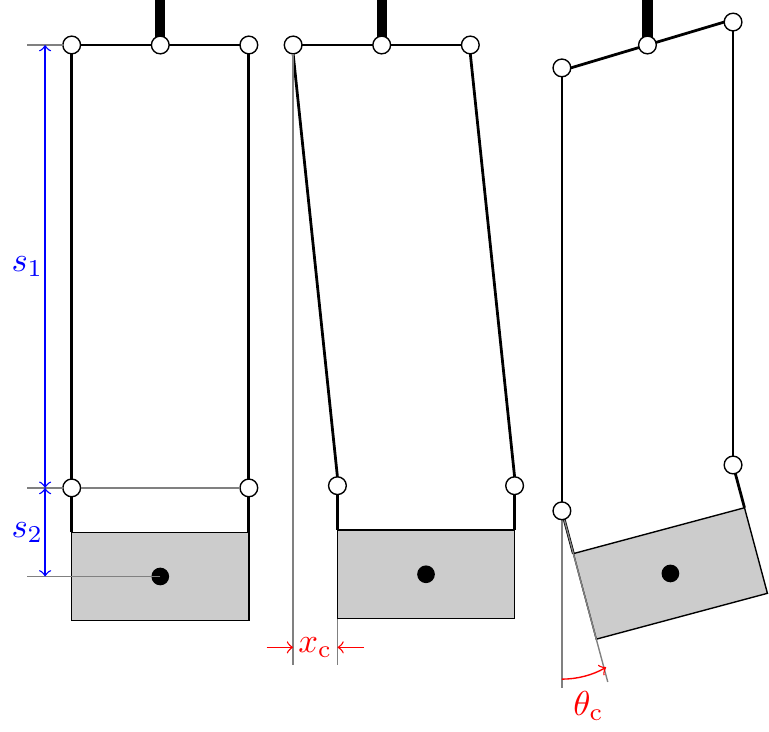}
\caption{A coil (grey box) with mass $M$ is suspended from an ideal linkage. The empty circles are flexible pivots. The  suspension rods have a length $s_1$. The vertical distance between the center of mass of the coil and the lower flexures is $s_2$. We assume very compliant flexures (no stiffness) and neglect the mass of the suspension. }
\label{fig:sec6:sus}
\end{figure}

To each of the stiffness, two factors contribute, the gravitational stiffness and the elastic stiffness of the flexure joints. The former is the dominating term and, hence,  for a back-of-the-envelope calculation, the elastic stiffness can be neglected. Figure~\ref{fig:sec6:sus} shows a toy model of a coil suspension. A coil of mass $M$ is suspended by rods of length $s_1$. The open circles in the figure denote ideal (zero stiffness) flexures. The vertical distance between the center of mass of the coil and the lower flexure plane is $s_2$. The suspension is assumed to be massless.

If the coil moves horizontally by $x_\cc$, the center of mass of the coil rises (to the second order) by
\begin{equation}
    \Delta z \approx \frac{x_\cc^2}{2 s_1}.
\end{equation}
Hence the gravitational energy of the coil increases by $E_\mathrm{grav} = Mg \Delta z$. Thus, the horizontal force and consequently the horizontal stiffness are given by
\begin{equation}
F_x =-\frac{\partial E_\mathrm{grav}}{\partial x_\cc}=-\frac{Mg}{s_1} x_\cc,
\end{equation}
and
\begin{equation}
k_x=-\frac{F_x}{x_\cc} = \frac{Mg}{s_1}.
\label{eq:sec6:k:guess}
\end{equation}

A rotation of the coil by the angle $\theta_\cc$ causes a vertical change in the center of mass that is approximately,
\begin{equation}
    \Delta z \approx \frac{\theta_\cc^2}{2} s_2.
\end{equation}
This equation leads to the torque and the rotational stiffness. It is,
\begin{equation}
\tau =-\frac{\partial E_\mathrm{grav}}{\partial \theta_\cc}=-Mg s_2 \theta_\cc 
\end{equation}
and
\begin{equation}
\kappa_\theta=-\frac{\tau}{\theta_\cc} = Mgs_2.
\label{eq:sec6:kappa:guess}
\end{equation}
In the above approximation, the ancillary variables $\xi$ and $\eta$ evaluate to
\begin{equation}
\xi = \ \sqrt{ s_2 s_1} \;\mbox{and}\;
\eta = M g \sqrt{\frac{\ds s_2}{\ds  s_1}}.
\end{equation}

With equations~(\ref{eq:sec6:k:guess}) and (\ref{eq:sec6:kappa:guess}), estimates for $k_x$ and $\kappa_\theta$ can be calculated. Typically, the mass of the coil is about 10 times that of the test mass. For a back-of-the-envelope calculation, $s_1$ can be estimated from the length of the suspension rods and $s_2$ from the coil height. Typical values are  $s_1\sim\SI{1}{\meter}$, and $s_2\sim \SI{1}{\centi \meter}$. The  measured values for the NIST-4 Kibble balance  can be found in table~\ref{tab:sec6:nist4:dims}.

The values for $k_x$ and $\kappa$ are not exact matches, but the order of magnitudes agrees. the geometric determination of $k_x$ is about 51\,\% higher or $1.8\,\sigma$. 
The value for $\kappa_\theta$ determined by the measured coil displacement is about twice as much as the value obtained from the geometry.  So, it appears that neglecting the elastics restoring component is not justified for the rotation  of the coil.

Although the agreement is not perfect, the values obtained from the measured coil position can provide useful information to the experimenter. Also, note that the stiffnesses are a byproduct of the analysis carried out here. The main results are $x_o$ and $\theta_o$.

\begin{table}[h!]
\begin{center}
\caption{The dimensions of the NIST-4 coil. The geometric stiffnesses are calculated from the first three rows. The measured values come from table~\ref{tab:sec6:results} }
\label{tab:sec6:nist4:dims}
\begin{tabular}{cS[table-format=3.2]l S[table-format=3.2(2)] l}
\hline
&\multicolumn{2}{c}{geometric}
&\multicolumn{2}{c}{measured} \\
quantity 
&\multicolumn{1}{c}{value}
&\multicolumn{1}{c}{unit} 
&\multicolumn{1}{c}{value}
&\multicolumn{1}{c}{unit} \\
\hline
$M$ & 8.6&\si{\kilo \gram} \\
$s_1$ & 87.3&\si{\centi\meter} \\
$s_2$ & 4.8&\si{\centi\meter} \\
$k_x$ &  96.6&\si{\newton \per \meter} &64(18)&\si{\newton \per \meter} \\
$\kappa_\theta$ & 4.0&\si{\newton \meter \per \radian}&8.0(6)\si{\newton \meter \per \radian}\\
$\xi$ &  0.21&\si{\meter}&0.35(22)&\si{\meter}\\
$\eta$ &  19.8&\si{\newton}&23(9)&\si{\newton}\\
\hline
\end{tabular}
\end{center}
\end{table}

\subsection{Final remarks}
By analyzing the coil position with the procedure outlined above, the absolute position and orientation of the coil with respect to the magnet can be obtained. Here, we have only discussed the 1-dimensional case, i.e., two unknowns $x_o$ and $\theta_o$. In  a real system, four parameters need to be determined. An offset along the $xy$-directions and rotations about the $x$ and $y$ axes. \ref{secb} gives the solution for this case. The good news is that the two directions are independent of each other. So, each direction can be analyzed with the algorithm described above. For one direction $x_\cc$ and $\theta_y$ are used. For the other one, the analysis is done with $y_\cc$ and $\theta_x$.

Once the position of the coil relative to the magnet is obtained and it is not satisfactory, the experimenter has to move either the Kibble balance or the magnet to minimize $x_o$ and $\theta_o$. For the article, we assumed that the magnet is vertical and the coil is tilted. Hence, $\theta_o$ indicates a coil tilt with respect to the vertical direction. If the magnet is not perfectly vertical, $\theta_o$ will be the angular difference between the coil and the magnet. 

This will lead to another interesting possibility. Researchers have spent significant effort to align the magnet to vertical in the laboratory, e.g.~\cite{bielsa2015alignment}. However, with the method shown here, the relative angle can be obtained, and it is relatively simple to measure the absolute angle of the coil in the laboratory. Hence, the absolute angle of the magnet can be obtained. With this information, the magnet can be leveled.

\section{Summary and Conclusions}
In this article, we have used the flux picture to derive interesting properties of electromagnetic torques acting on Kibble balance coils. The magnetic flux through the inner yoke is of prime importance for yoke-based permanent magnet systems. This flux can easily be calculated using the pillbox method. It relies on the fact that the magnetic flux through all surfaces  enclosing  a volume sums to zero -- a consequence of Maxwell's equation $\vec\nabla \cdot \vec{B}=0$.

With a properly chosen pillbox, we showed that the torque on a coil depends on the derivative of the geometric factor, $\partial B_r/\partial z$. Most importantly, the torque vanishes if the current-carrying coil is at a local extremum, i.e.,  $\partial B_r/\partial z=0$.  This fact provides another reason to strive for a flat profile when designing a Kibble balance.

We then analyzed the torque that can arise when the coil is not centered in the magnet. The tilt and horizontal displacement lead to coupled equations because the flux through the coil contains the product of displacement and tilt angle. The coupled equations were solved, and a simple expression of the position of the coil with current as a function of the coil misalignment was found. From that solution, we found a recipe that yields the coil misalignment from the measured coil position with no, positive, and negative current. This result can be used to align the coil and even the magnet vertically in the laboratory.

\section*{Acknowledgements}
The authors would like to thank the anonymous reviewers for carefully reading the text and providing valuable suggestions to improve the manuscript.

\appendix

\section{About signs}
\label{seca}
\begin{figure}
\centering
\includegraphics[width=\columnwidth]{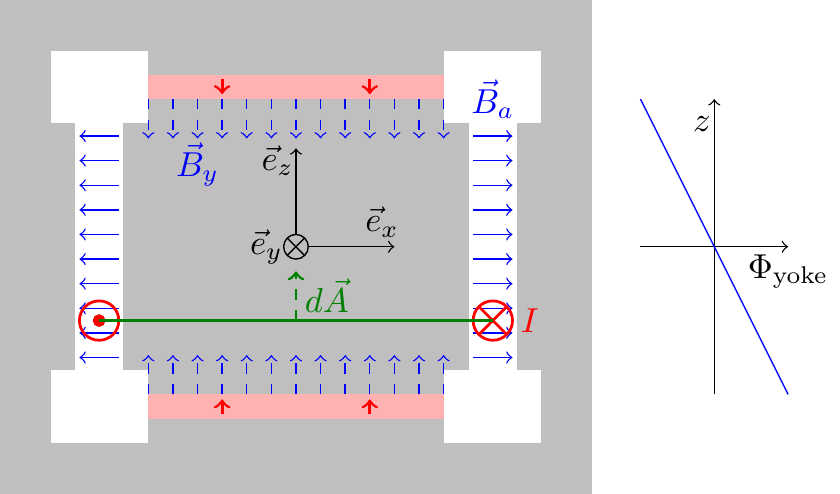}
\caption{A toy model of a current-carrying coil in a BIPM magnet. The red circles are cross-sectional views of the coil. The current flows into the paper plane on the right side. The solid blue arrows denote the flux density in the air gap. The dashed blue arrows the flux density in the yoke. The green horizontal line is the area of the coil, with the differential area vector $\ud\vec{A}$ pointing vertically. The graph on the right-hand side shows the flux through the yoke. The flux gradient is negative.}
\label{fig:mag:signs}
\end{figure}

The purpose of this section is to clarify the sign of the flux and the sign in front of the flux derivative for the force and torque calculations. Figure~\ref{fig:mag:signs} shows a toy model of a current-carrying coil in a magnet system. We calculate the direction of the force in two ways, first with the Lorentz force and then with the flux integral. 

The Lorentz force for a piece of wire in a magnetic field with flux density (in the air gap) $\vec{B}_\mathrm{a}$ is given by
\begin{equation}
    \vec{F} = q \vec{v} \times \vec{B}_\mathrm{a},
\end{equation}
where $q\vec{v}$ points in the direction of the current. We consider the piece of wire of length $d\ell$ that is shown on the right side of the gap. Here $q \vec{v}$ points in the $y$ direction and $\vec{B_a}$ in the $x$ direction. Hence the magnetic force points in $\vec{e}_y\times \vec{e}_x$ which equals to $-\vec{e}_z$. It can easily be seen that this direction is true for the complete coil by integrating around the loop. So, the force on the coil in the toy model is downward.

To calculate the force using the flux integral, we need to calculate the energy of the system,
\begin{equation}
    E = -NI \int_A \vec{B}_\mathrm{y} \ud\vec{A}.
    \label{eq:app:energy}
\end{equation}
The integral needs to be carried out over the open area of the coil. But, as has been discussed in section~\ref{sec2}, in a symmetric case, it is sufficient only to take the yoke into consideration because the field in the air gap is perpendicular to the area vector and does not contribute. So, $A$ is the cross-sectional area of the inner yoke, and  $B_y$ is the magnetic flux density in the inner yoke. 
We need to find the direction of $\ud \vec{A}$. The convention is that the  $\ud \vec{A}$ can be found by the right-hand rule. The fingers, except for the thumb, point in the direction of the current, and the thumb points in the direction of $\ud \vec{A}$. For the toy model, $\ud \vec{A}$ points up, parallel with $\vec{e}_z$.  In this situation, the flux integral ($\int_A \vec{B}_\mathrm{y} \ud\vec{A}$) is negative with the coil at the top of the gap and positive at the bottom. The electromagnetic  energy of the system is
\begin{equation}
    E = NI B(r_\mathrm{i}) 2\pi r_\mathrm{i} z_\mathrm{c},
\end{equation}
where $r_\mathrm{i}$ is the radius of the inner yoke.
See the text above equation~(\ref{eq:sec2:flux_yoke_BIPM}) for a derivation of the exact expression of the flux integral. Here only the sign is important. The force is given by the negative derivative of the observed coordinate, here
\begin{equation}
    F_\mathrm{coil} = -\frac{\partial E}{\partial z_\mathrm{c}} = -NI B(r_\mathrm{i}2\pi r_\mathrm{i}.
\end{equation}
The negative derivative of the energy (area integral) and the line integral produce a consistent direction of the force. In both cases, the force points in the negative $z$ direction.

With the flux integral, the force on the magnet can also be calculated if the magnet is displaced by $z_m$ as is the case in the UME (Ulusal Metroloji Enstitüsü, Turkey)  watt balance~\cite{UME}, the energy is given by
\begin{equation}
    E = NI B(r_\mathrm{i} 2\pi r_\mathrm{i} (z_\mathrm{c}-z_\mathrm{m}),
\end{equation}
and, hence,
\begin{equation}
F_\mathrm{magnet}= -\frac{\partial E}{\partial z_\mathrm{m}} = NI B(r_\mathrm{i})2\pi r_\mathrm{i} = - F_\mathrm{coil}
\end{equation}
As one would expect from Newton's third law, the force on the magnet, $F_\mathrm{magnet}$  is equal and opposite to the force on the coil, $F_\mathrm{c}$. And, of course, that is true whether the magnet is moved or not.

The last remaining sign that needs to be discussed is why there is a minus sign in equation~\ref{eq:app:energy}.

This negative sign can be obtained by considering the coil as a dipole. The dipole moment of a current-carrying  wire loop with $N$ turns  is
\begin{equation}
    \vec{m} = NI\vec{A},
\end{equation}
where the direction of the area vector is obtained by the right-hand rule. A dipole in a magnetic flux density has the lowest energy when $\vec{m}$ and $\vec{B}_y$ are parallel. Thus, the energy of the dipole in flux density $\vec{B}_y$ is
\begin{equation}
    E=-\vec{m}\cdot\vec{B}_\mathrm{y} = -NI\vec{A}\cdot\vec{B}_\mathrm{y}=-NI \int \vec{B}_\mathrm{y} \ud\vec{A}.
\end{equation}
Equation~\ref{eq:app:energy} is obtained with the correct sign.

\section{Coil tilt and displacement in two dimensions}
\label{secb}
In the main text, the energy of a coil that is displaced and tilted is calculated. For simplicity, only the case where the coil was rotated about the $y$-axis and displaced along the $x$-axis was discussed. Here, we calculate the general case. The coil is tilted by a small angle about any axis in the $xy$ plane and is also displaced along any direction in that plane.

\begin{figure}
\centering
\includegraphics[width=\columnwidth]{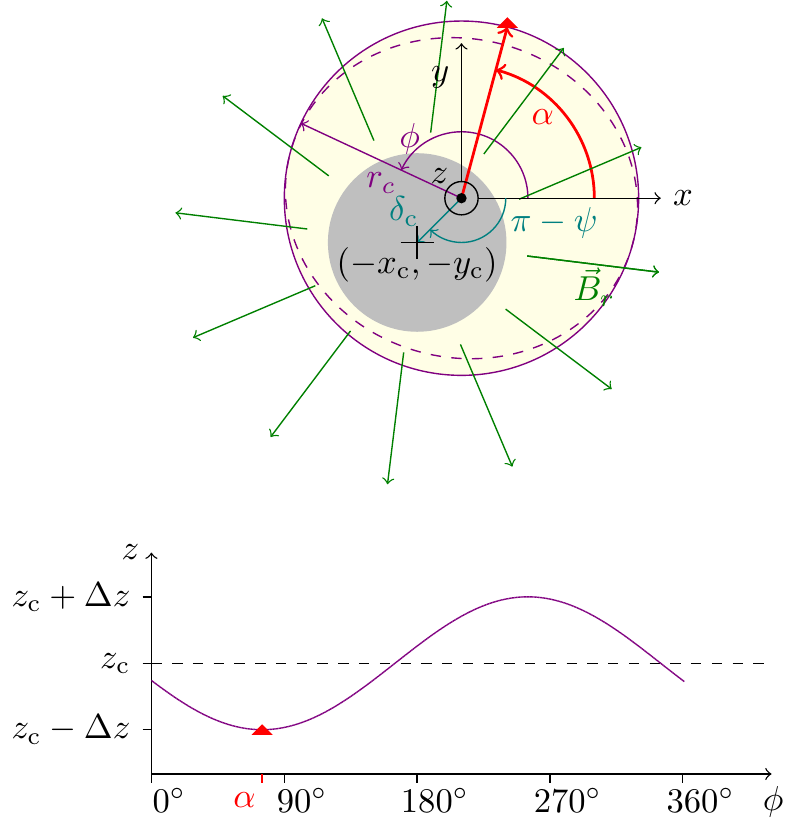}
\caption{Similar drawing to the lower part of figure~\ref{fig:sec4:tilted:displaced:pillbox}. But here, the coil is translated in $x$ and $y$ with respect to the symmetry axis of the yoke by $x_\cc$ and $y_\cc$. Since the coordinate system is chosen to be centered on the coil, the center of the yoke is at $(-x_\cc,-y_\cc)$. In figure~\ref{fig:sec4:tilted:displaced:pillbox} the lowest point of the coil was along the $x$-axis, i.e., the coil was rotated about the $y$-axis. Here the lowest point can occur at an arbitrary angle $\alpha$, which is marked by a triangle. The lower graph shows the outer edge of the top surface of the pillbox. The triangle marks the lowest point, again.
\label{fig:secb:topview}}
\end{figure}

The coordinates of the coil in a coordinate system centered on the symmetry line of the yoke are given by
\begin{equation}
    x_\cc = -\delta_\cc \cos\psi\;\mbox{ and} \;y_\cc = -\delta_\cc \sin\psi,
\end{equation}
where $\delta_\cc$ is the  absolute distance between the two.
Here, we use a coordinate system that is centered on the coil, see figure~\ref{fig:secb:topview}. Hence, the center of the yoke is at $(-x_\cc,-y_\cc)$.

Then, equation~(\ref{eq:sec4:Br}) must be modified to read
\begin{multline}
\vec{B}(r)\ud \phi = \frac{B_r(r_\mathrm{i}) r_\mathrm{i}} {( x_\cc+r_\cc\cos\phi)^2+(y_\cc+r_\cc\sin\phi)^2} \times\\
    \left(
    \begin{array}{l}
    x_\cc+r_\cc \cos\phi \\
    y_\cc+r_\cc \sin\phi\\
    0
    \end{array}
    \right).
    \label{eq:secb:Br}
\end{multline}

To account for the rotation about a different axis, the height of the pillbox, equation~(\ref{eq:sec3:height}) must be changed to
\begin{equation}
    z_t(\phi,\theta_\cc) = z_\cc-r_\cc\tan\theta_\cc \cos{(\phi-\alpha)}.
    \label{eq:secb:height}
\end{equation}
With these two new parametrizations, the integral in equation~(\ref{eq:sec4:integral}) yields
\begin{equation}
\Phi_\mathrm{s} =B_r \pi r_\mathrm{i} (2 z_\cc + \delta_\cc \cos{(\alpha -\psi)} \tan\theta_\cc.
\label{eq:secB:integral:result}
\end{equation}
Using $\Phi_\mathrm{t} =-\Phi_\mathrm{s}+\Phi_\mathrm{b}$, $E=NI\Phi_\mathrm{t}$, and $F_z = -\partial E/\partial z_\cc$ yields
\begin{equation}
E = -F_z \left(z_\cc + \frac{\delta_\cc}{2} \cos{\left(\alpha -\psi\right)} \tan\theta_\cc\right) +NI\Phi_\mathrm{b},
\label{eq:secB:energy1}
\end{equation}
where $NI\Phi_\mathrm{b}$ denotes an inconsequential constant.

To calculate the torques on the coil about the $x$ and $y$ axes, we need to digress into rotation matrices in $\Re^3$. The rotation about the $x$ axis with an angle $\theta_x$ is given by
\begin{equation}
    \mathbf{R}_x = \left(
    \begin{array}{ccc}
    1 &0& 0\\
    0 &\cos\theta_x &-\sin\theta_x\\
    0 &\sin\theta_x &\cos\theta_x
    \end{array}
    \right).
\end{equation}
Similarly, the rotation about the $y$-axis by $\theta_y$  can be written as
\begin{equation}
    \mathbf{R}_y = \left(
    \begin{array}{ccc}
    \cos\theta_y&0 &\sin\theta_y\\
    0 &1& 0\\
    -\sin\theta_y&0 &\cos\theta_y
    \end{array}
    \right).
\end{equation}
In general, the multiplication of the rotation matrices does not commute, i.e.,
\begin{equation}
    \mathbf{R}_x  \mathbf{R}_y \ne \mathbf{R}_y  \mathbf{R}_x.
\end{equation}
For very small angles, $\theta_x<<1$ and $\theta_y<<1$  and to  first order in the angles, the multiplication of the rotation matrices commutes, and the product is
\begin{equation}
\mathbf{R}_{xy} = \left(
    \begin{array}{ccc}
    1&0 &\theta_y\\
    0 &1& -\theta_x\\
    -\theta_y&\theta_x &1
    \end{array}
    \right).
\end{equation}
If $\vec{A}=(0,0,A)^T$ is the area vector of the horizontal coil, after rotation, it is,
\begin{equation}
\mathbf{R}_{xy}\vec{A} = A
\left(
    \begin{array}{c}
    \theta_y\\
    -\theta_x\\
    1
    \end{array}
    \right).
\end{equation}
Projecting $\mathbf{R}_{xy}\vec{A}$ down to the $xy-$plane gives the coordinates $(A\theta_y,-A\theta_x)$. This projection can be written in polar coordinates as  $(A\theta_\cc \cos\alpha,A\theta_\cc \sin\alpha)$, with
\begin{equation}
\tan\alpha = \frac{-\theta_x}{\theta_y}.
\end{equation}
and
\begin{equation}
\theta_\cc^2 = \theta_x^2+\theta_y^2.
\end{equation}
Hence, for small angles $\theta_\cc$ ($\tan\theta_\cc\approx\theta_\cc$),
\begin{equation}
\theta_y = \theta_\cc \cos\alpha\;\;\mbox{and}\;\;\theta_x = -\theta_\cc \sin \alpha.
\end{equation}
With that, equation~(\ref{eq:secB:energy1}) can be written as a function of the five coil coordinates. It is
\begin{equation}
E = -F_z z_\cc + \frac{F_z y_\cc \theta_x}{2} - \frac{F_z x_\cc \theta_y}{2}  +NI\Phi_\mathrm{b},
\label{eq:secB:energy2}
\end{equation}

Following equation~(\ref{eq:sec5:energy}), the total energy of the coil, including the mechanical and magnetic energy of the coil, can be calculated. It is 
\begin{multline}
    E = -F_z z_\cc+NI\Phi_\mathrm{b} + \frac{F_z y_\cc \theta_x}{2} - \frac{F_z x_\cc \theta_y}{2} \\
+ \frac{1}{2} k_{xy} (x_\mathrm{c}-x_o)^2 
+ \frac{1}{2} k_{xy} (y_\mathrm{c}-y_o)^2  \\
+ \frac{1}{2} \kappa_\theta (\theta_\mathrm{x}-\theta_{xo})^2 
+ \frac{1}{2} \kappa_\theta (\theta_\mathrm{y}-\theta_{yo})^2.
\label{eq:secb:energy}
\end{multline}
It is assumed the stiffness for translation of the coil along $x$ is the same as along $y$, $k_{xy}$. Furthermore, the rotational stiffnesses for rotating about the $x$ and $y$ axis are identical. They are $\kappa_\theta$. Similar to the main text, in equation~(\ref{sec5:deriv:eq:zero}) the derivatives of the energy with respect to the coil parameters can be set to zero and a matrix equation is obtained, see equation~(\ref{sec5:matrix}). Here the energy derivatives with respect to $x_\cc$,$y_\cc$,$\theta_x$, and $\theta_y$ are calculated. The matrix equation is
\begin{multline}
\left(
\begin{array}{cccc} 
k_{xy} & 0 &0 & -F_z/2 \\ 
0  & k_{xy} & F_z/2 & 0 \\
0 & F_z/2 &\kappa_\theta & 0\\
- F_z/2&0  &0&\kappa_\theta 
\end{array}\right)
\left(\begin{array}{c} 
x_\mathrm{c} \\ 
y_\mathrm{c} \\ 
\theta_x\\
\theta_y
\end{array}\right) =\\
\left(\begin{array}{c} 
k_{xy} x_o \\ 
k_{xy} y_o \\ 
\kappa_\theta \theta_{xo}\\
\kappa_\theta \theta_{yo}
\end{array}\right).
\label{secb:matrix}
\end{multline}
The solutions are
\begin{multline}
\vec{p}_\cc=
\left(
\begin{array}{l} 
x_\cc\\
y_\cc\\
\theta_x\\
\theta_y
\end{array}\right) =
\frac{2}{4k_x\kappa_\theta -F_z^2}
 \times \\
\left(\begin{array}{l} 
2 k_{xy}\kappa x_o+\kappa_\theta\theta_{yo} F_z\\ 
2 k_{xy}\kappa y_o-\kappa_\theta\theta_{xo} F_z\\ 
-k_{xy} y_o F_z+2k_{xy}\kappa_\theta \theta_{xo} \\ 
k_{xy} x_o F_z+2k_{xy}\kappa_\theta\theta_{yo} \\
\end{array}\right) 
\end{multline}

Similar to equation~(\ref{eq:sec5:displacement:matrix}) in the main text, a displacement matrix can be introduced. It is,
\begin{multline}
\mathbf{D} =
\frac{2}{4k_x\kappa_\theta -F_z^2} \times\\
\left(\begin{array}{llll}
2k_{xy} \kappa_\theta & 0&0&  \kappa_\theta  F_z \\ 
0&2k_{xy} \kappa_\theta &   -\kappa_\theta  F_z & 0\\ 
0&-k_{xy} F_z &   2k_{xy}\kappa_\theta  & 0\\ 
k_{xy} F_z &0 &0 & 2k_{xy}\kappa_\theta 
\end{array}\right) .
\label{eq:secb:displacement:matrix}
\end{multline}
With the displacement matrix, it is
\begin{equation}
\vec{p}_\cc= \mathrm{D} \vec{p}_o\;\mbox{with}\;
\vec{p}_o=
\left(\begin{array}{c} 
x_o \\ 
y_o \\ 
\theta_{xo}\\
\theta_{yo}
\end{array}\right).
\end{equation}
Equations~(\ref{eq:sec5:matrix:sol}) and (\ref{eq:sec5:displacement:matrix}) are a subsection of the above.

\section{Torque on a displaced coil using the line integral}
\label{secc}
The point of this article is to encourage the use of the area integral instead of the line integral to calculate forces and torques on the coil. Here, we would like to verify the torque on the simply displaced coil, equation~(\ref{eq:sec4:tau:1}) using the line integral.

\begin{figure}
\centering
\includegraphics[width=0.8\columnwidth]{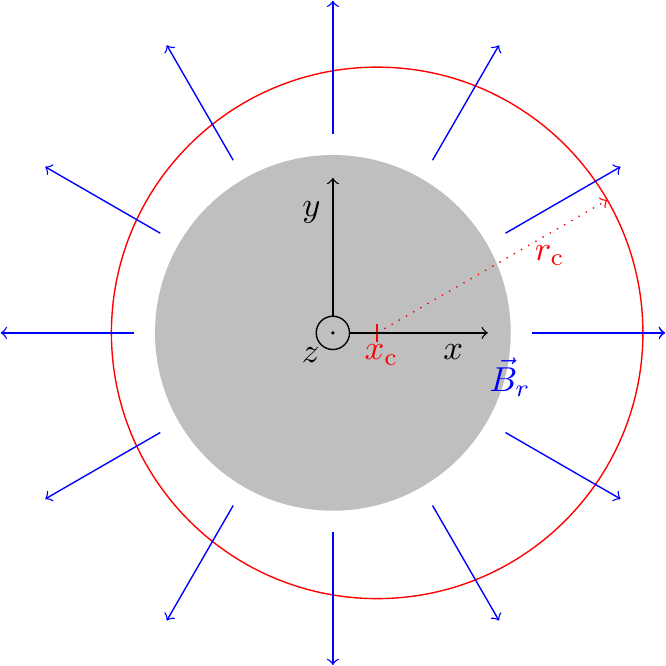}
\caption{A coil with radius $r_\mathrm{c}$ is translated by $x_\mathrm{c}$ from the origin centered on the inner yoke shown in grey}
\label{fig:secc:coil}
\end{figure}

Figure~\ref{fig:secc:coil} shows the scenario that is calculated below. A coil with radius $r_\mathrm{c}$ is displace by the $x_\mathrm{c}$ from the origin. The field is azimuthally symmetric and is only vertical. Hence the radial field component at the location $\vec{r}=r\vec{e}_r$ is given by
\begin{equation}
\vec{B} = \vec{e}_r B(r_\mathrm{i}) \frac{r_\mathrm{i}}{r},
\end{equation}
The position of the wire  in Cartesian coordinates is given by
\begin{equation}
\vec{\ell} = \vec{\ell}_\mathrm{c}+\left(
\begin{array}{l}
r_\mathrm{c}  \cos\phi\\
r_\mathrm{c}  \sin\phi\\
0
\end{array}
\right),
\end{equation}
where $\vec{\ell}_\mathrm{c} = (x_\mathrm{c},0,0)^T$
Hence, the line element is
\begin{equation}
\ud\vec{\ell} = \ud\phi
\left(
\begin{array}{l}
-r_\mathrm{c}  \sin\phi\\
r_\mathrm{c}  \cos\phi\\
0
\end{array}
\right) 
\end{equation}
Here we calculate the torque on the coil with respect to $\vec{\ell}_\mathrm{c}$. It is
\begin{equation}
\vec{N} =  I \oint_C  \left(\vec{\ell}-\vec{\ell}_\mathrm{c}\right) \times (\vec{B}(|\vec{\ell}|)\times \ud \vec{\ell})
\label{eq:secc:int}
\end{equation}
The term $(\vec{B}(|\vec{\ell}|)\times \ud \vec{\ell})$ only has a vertical component. It is
\begin{equation}
(\vec{B}(|\vec{\ell}|)\times \ud \vec{\ell})_z=
\frac{B(r_\mathrm{i}) r_\mathrm{i} r_\cc \ud \phi}{\sqrt{x_\mathrm{c}^2+r_\mathrm{c}^2+2x_\mathrm{c} r_\mathrm{c}\cos\phi}}
\end{equation}
Integrating equation~(\ref{eq:secc:int}) yields $N_y=0$ and 
\begin{equation}
N_x\approx I  B(r_\mathrm{i}) r_\mathrm{i} \pi x_\mathrm{c}= \frac{F_z}{2}  x_\mathrm{c}.
\end{equation}
The above equation agrees with equation~(\ref{eq:sec4:tau:1}).

\section*{References}
\providecommand{\newblock}{}


\begin{thebibliography}{10}
\expandafter\ifx\csname url\endcsname\relax
  \def\url#1{{\tt #1}}\fi
\expandafter\ifx\csname urlprefix\endcsname\relax\def\urlprefix{URL }\fi
\providecommand{\eprint}[2][]{\url{#2}}

\bibitem{Robinson16}
Robinson I~A and Schlamminger S 2016 {\em Metrologia\/} {\bf 53} A46--A74
  \urlprefix\url{https://doi.org/10.1088/0026-1394/53/5/A46}

\bibitem{Bartl_2017}
Bartl G {\em et~al.\/} 2017 {\em Metrologia\/} {\bf 54} 693--715
  \urlprefix\url{https://doi.org/10.1088/1681-7575/aa7820}

\bibitem{Stock2020}
Stock M {\em et~al.\/} 2020 {\em Metrologia\/} {\bf 57} 07030
  \urlprefix\url{https://doi.org/10.1088/0026-1394/57/1a/07030}

\bibitem{haddad2016bridging}
Haddad D, Seifert F, Chao L~S, Li S, Newell D~B, Pratt J~R, Williams C and
  Schlamminger S 2016 {\em Metrologia\/} {\bf 53} A83--A85
  \urlprefix\url{https://doi.org/10.1088/0026-1394/53/5/A83}

\bibitem{Sasso2014}
Sasso C~P, Massa E and Mana G 2014 {\em Metrologia\/} {\bf 51} S65--S71
  \urlprefix\url{https://doi.org/10.1088/0026-1394/51/2/s65}

\bibitem{Shisong2022}
Li S and Schlamminger S 2022 {\em Metrologia\/} {\bf 59} 022001
  \urlprefix\url{https://doi.org/10.1088/1681-7575/ac464a}

\bibitem{li2016coil}
Li S, Schlamminger S, Haddad D, Seifert F, Chao L and Pratt J~R 2016 {\em
  Metrologia\/} {\bf 53} 817
  \urlprefix\url{https://doi.org/10.1088/0026-1394/53/2/817}

\bibitem{Kibble1976}
Kibble B~P 1976 A measurement of the gyromagnetic ratio of the proton by the
  strong field method {\em Atomic masses and fundamental constants 5\/}
  (Springer) pp 545--551

\bibitem{NPL}
Robinson I~A 2012 {\em Metrologia\/} {\bf 49} 113--156
  \urlprefix\url{https://doi.org/10.1088/0026-1394/49/1/016}

\bibitem{nist3}
Schlamminger S, Haddad D, Seifert F, Chao L~S, Newell D~B, Liu R, Steiner R~L
  and Pratt J~R 2014 {\em Metrologia\/} {\bf 51} S15--S24
  \urlprefix\url{https://doi.org/10.1088/0026-1394/51/2/S15}

\bibitem{BIPMmag2006}
Stock M 2006 {\em INFOSIM Inform. Bull. Inter Amer. Metrol. Syst.\/} {\bf 9}
  9--13

\bibitem{LNEmag}
Gournay P, Genev{\`e}s G, Alves F, Besbes M, Villar F and David J 2005 {\em
  IEEE Trans. Instrum. Meas.\/} {\bf 54} 742--745
  \urlprefix\url{https://doi.org/10.1109/TIM.2004.843072}

\bibitem{MSL}
Sutton C~M and Clarkson M~T 2014 {\em Metrologia\/} {\bf 51} S101--S106
  \urlprefix\url{https://doi.org/10.1088/0026-1394/51/2/S101}

\bibitem{NISTmag}
Seifert F, Panna A, Li S, Han B, Chao L, Cao A, Haddad D, Choi H, Haley L and
  Schlamminger S 2014 {\em IEEE Trans. Instrum. Meas.\/} {\bf 63} 3027--3038
  \urlprefix\url{https://doi.org/10.1109/TIM.2014.2323138}

\bibitem{li2021resolution}
Li S, Schlamminger S, Marangoni R, Wang Q, Haddad D, Seifert F, Chao L, Newell
  D and Zhao W 2021 {\em Scientific reports\/} {\bf 11} 1--13

\bibitem{li17}
Li S, Bielsa F, Stock M, Kiss A and Fang H 2017 {\em Metrologia\/} {\bf 55}
  75--83

\bibitem{NPL3}
Robinson I~A, Berry J, Bull C, Davidson S, Jarvis C, Lovelock P, Lucas C,
  Urquhart J, Webster E and Williams P 2018 Developing the next generation of
  \textsc{NPL} \textsc{K}ibble balances {\em 2018 Conference on Precision
  Electromagnetic Measurements (CPEM 2018)\/} (IEEE) pp 1--2

\bibitem{hysteresis}
Li S, Bielsa F, Stock M, Kiss A and Fang H 2020 {\em IEEE Transactions on
  Instrumentation and Measurement\/} {\bf 69} 5717--5726
  \urlprefix\url{http://doi.org/10.1109/TIM.2019.2962848}

\bibitem{Kibble_2014}
Kibble B~P and Robinson I~A 2014 {\em Metrologia\/} {\bf 51} S132--S139
  \urlprefix\url{https://doi.org/10.1088/0026-1394/51/2/s132}

\bibitem{haddad2016invited}
Haddad D, Seifert F, Chao L, Li S, Newell D, Pratt J, Williams C and
  Schlamminger S 2016 {\em Review of Scientific Instruments\/} {\bf 87} 061301
  \urlprefix\url{https://doi.org/10.1063/1.4953825}

\bibitem{bielsa2015alignment}
Bielsa F, Lu Y, Lavergne T, Kiss A, Fang H and Stock M 2015 {\em Metrologia\/}
  {\bf 52} 775 \urlprefix\url{https://doi.org/10.1088/0026-1394/52/6/775}

\bibitem{UME}
Ahmedov H, A{\c{s}}k{\i}n N~B, Korutlu B and Orhan R 2018 {\em Metrologia\/}
  {\bf 55} 326--333 \urlprefix\url{https://doi.org/10.1088/1681-7575/aab23d}

\end{thebibliography}
\end{document}